\documentclass{article}
\usepackage[a4paper,
            bindingoffset=0.2in,
            left=0.6in,
            right=0.6in,
            top=1in,
            bottom=1in,
            footskip=.25in]{geometry}
\usepackage[utf8]{inputenc}
\usepackage{jcappub}
\usepackage{natbib}
\setcitestyle{square,numbers}
\usepackage{amsmath}
\usepackage{bbold}
\usepackage{booktabs}
\usepackage{graphicx}
\usepackage{floatpag}
\usepackage{xcolor}
\usepackage{float}
\usepackage{multirow,array,longtable}
\usepackage{arydshln} 
\usepackage[makeroom]{cancel}
\usepackage{amsfonts}
\usepackage{stackrel}
\usepackage{slashed}

\renewcommand\eqref[1]{Eq.~(\ref{#1})}
\newcommand\eqrefs[2]{Eqs.~(\ref{#1})-(\ref{#2})}
\newcommand\figref[1]{Fig.~\ref{#1}}
\newcommand\figrefs[2]{Figs.~\ref{#1}-\ref{#2}}
\newcommand\tabref[1]{Table~\ref{#1}}

\newcommand\secref[1]{Section~\ref{#1}}
\newcommand\appref[1]{Appendix~\ref{#1}}
\newcommand{\be}{\begin{equation}}
\newcommand{\ee}{\end{equation}}
\newcommand{\bear}{\begin{eqnarray}}
\newcommand{\eear}{\end{eqnarray}}
\newcommand{\nn}{\nonumber}

\newcommand{\mL}{\mathcal{L}}
\newcommand{\mO}{\mathcal{O}}
\newcommand{\mM}{\mathcal{M}}

\newcommand{\mS}{\mathcal{S}}
\newcommand{\mC}{\mathcal{C}}
\newcommand{\mF}{\mathcal{F}}

\newcommand{\mB}{\mathcal{B}}

\def\Tr{{\rm Tr}}
\def\vev{v}

\def\sw{s_{\rm w}}

\def\mh{m_{\rm H}}
\def\mw{m_{\rm W}}
\def\mz{m_{\rm Z}}
\def\wz{\Gamma_{\rm Z}}
\def\mtop{m_{\rm top}}

\def\kCP{\kappa_{\rm CP}^l}
\def\dCP{\delta_{\rm CP}^l}
\def\cCP{\cos\dCP}
\def\sCP{\sin\dCP}
\def\cosCP{c_{\rm CP}}
\def\sinCP{s_{\rm CP}}
\def\Ctree{A_{\rm Tree}}
\def\dilep{l\bar{l}}

\def\mlep{m_l}
\def\ylep{Y_l}
\def\betal{\beta_l}
\def\thlep{\theta_{\gamma l}}
\def\clep{\cos(\thlep)}
\def\slep{\sin(\thlep)}
\def\costhlep{c_{\gamma l}}
\def\sinthlep{s_{\gamma l}}
\def\ditau{\tau\bar{\tau}}
\def\mditau{m_{\ditau}}
\def\mtau{m_\tau}
\def\thtau{\theta_{\gamma\tau}}
\def\ctau{\cos(\thtau)}
\def\dimu{\mu\bar{\mu}}
\def\mdimu{m_{\dimu}}
\def\mmu{m_\mu}
\def\thmu{\theta_{\gamma\mu}}
\def\cmu{\cos(\thmu)}
\def\diel{e\bar{e}}
\def\mdiel{m_{\diel}}
\def\mel{m_e}
\def\thel{\theta_{\gamma e}}
\def\cel{\cos(\thel)}
\def\Mertri{\mM_3}
\def\aGTE{\mF_3}
\def\CAtollbar{\mC_{l\bar{l}|\gamma}}
\def\CltoAlbar{\mC_{\gamma\bar{l}|l}}
\def\ClbartoAl{\mC_{\gamma l|\bar{l}}}
\def\Cllbar{\mC_{l\bar{l}}}
\def\CAlbar{\mC_{\gamma\bar{l}}}
\def\CAl{\mC_{\gamma l}}
\def\CAtotautau{\mC_{\tau\bar{\tau}|\gamma}}
\def\Ctautau{\mC_{\tau\bar{\tau}}}
\def\aone{a_1}
\def\atwo{a_2}
\def\aboth{a_{1,2}}
\def\bone{b_1}
\def\btwo{b_2}
\def\bboth{b_{1,2}}

\title{Tripartite entanglement and Bell non-locality in loop-induced Higgs boson decays}
\author[a]{R. A.  Morales}
\affiliation[a]{IFLP, CONICET - Dpto. de Física, Universidad Nacional de La Plata, \\ 
C.C. 67, 1900 La Plata, Argentina}
\emailAdd{roberto.morales@fisica.unlp.edu.ar}

\abstract{
In this article, we study quantum entanglement properties of the three-body $H\to\gamma\dilep$ decays (for $l=e,\mu,\tau$) within the context of the Standard Model augmented with CP-violating interactions in the lepton Yukawa sector. 
Our aim is to elucidate the distribution of entanglement between the final photon, lepton and antilepton across the phase-space. These rare Higgs boson decays occur at 1-loop level, presenting a unique opportunity to scrutinize quantum correlations of fundamental interactions in tripartite systems by computing concurrence measures and investigating Bell non-locality. Moreover, we explore post-decay and autodistillation phenomena. 
Multipartite entanglement measures have much richer structure than those in the bipartite case, thus deserve more attention in collider phenomenology. In this line, we analyze here novel observables for these three-body Higgs boson decays, which can be extended to other multiparticle systems within the high-energy regime.
We found that entanglement manifests among final particles, occasionally achieving a maximally entangled state in specific kinematical configurations. Also, these decay channels are promising for Bell non-locality tests but CP-effects are suppressed by lepton masses in this kind of observables.
}

\begin{document}

\maketitle

\section{Introduction}
\label{section-intro}

Quantum entanglement stands as a pivotal resource for tasks which can not be performed via classical resources. Quantum Information theory develops the manipulation, control and distribution of the entanglement in a given system, with applications ranging from cryptography~\cite{PhysRevLett.67.661} to teleportation~\cite{PhysRevLett.70.1895} and quantum computation~\cite{PhysRevLett.86.5188}.
In particular, entanglement can be generated when two systems interact and elementary particle collisions, described by Quantum Field Theory (QFT), provide a natural framework for studying such properties of fundamental interactions.
However, it receives a very recent attention in the high-energy physics (HEP) community, see for instance review~\cite{Barr:2024djo} and references therein. 
Notably, the ATLAS Collaboration observed entanglement in $t\bar{t}$ production~\cite{Afik:2020onf,Afik:2022kwm,Fabbrichesi:2021npl,Severi:2021cnj,Aguilar-Saavedra:2022uye,Afik:2022dgh,Dong:2023xiw} with a significance exceeding 5$\sigma$~\cite{ATLAS:2023fsd}, despite collider detectors were not initially designed for probing such properties.

The $S$-matrix formulation in QFT allows to compute decay and scattering amplitudes at a given order in perturbation theory. Radiative corrections, arising from closed loops in the propagation of virtual particles, represent genuine quantum effects and the inhered correlations between initial and final states warrant attention.

In this work, we are interested on entanglement between the helicity degrees of freedom of the particles in the tripartite final state corresponding to the Higgs boson decaying into a photon and a lepton-pair. 
Previously, entanglement properties of bipartite Higgs decays were analyzed for the 2-qubit final states~\cite{Fabbrichesi:2022ovb,Altakach:2022ywa,Ma:2023yvd}, corresponding to tau-lepton pair and two photons. Higgs boson decaying into massive gauge bosons, i.e. 2-qutrits systems, were studied in~\cite{Barr:2021zcp,Aguilar-Saavedra:2022mpg,Aguilar-Saavedra:2022wam,Ashby-Pickering:2022umy,Fabbrichesi:2023cev,Bernal:2023ruk,Fabbri:2023ncz,Subba:2024aut}. Also, 2-qutrits systems were explored in diboson production at LHC~\cite{Fabbrichesi:2023cev,Ashby-Pickering:2022umy,Barr:2022wyq,Aoude:2023hxv} and through vector boson scattering~\cite{Morales:2023gow}.

On the other hand, tripartite entanglement within the HEP context is in development: Ref.~\cite{Acin:2000cs} explores positronium decaying into three photons, QED scattering processes with spectator particle were analyzed in~\cite{Fonseca:2021uhd,Quinta:2023ink,Blasone:2024dud}, Ref.~\cite{Sakurai:2023nsc} studies heavy fermion decaying into three fermions via generic (pseudo)scalar, (pseudo)vector and (pseudo)tensor interactions, Ref.~\cite{Konwar:2024nrd} explores three-flavor entanglement in neutrino oscillations, Ref.~\cite{Subba:2024mnl} worked in $t\bar{t}Z$ system at the LHC, and Ref.~\cite{Aguilar-Saavedra:2024whi} addresses entanglement among the two spins and total angular momentum in $H\to ZZ,WW$.
General properties of multipartite systems were presented in~\cite{Bernal:2023xzp,Bernal:2023jba}.
Furthermore, the tripartite Higgs boson decays considered here can be related to bipartite ones, in order to investigate post-decay entanglement~\cite{Aguilar-Saavedra:2023hss} and autodistillation phenomena~\cite{Aguilar-Saavedra:2024fig,Aguilar-Saavedra:2024hwd}.

Beyond the entanglement due to correlations among constituents of a system, is the concept of non-locality.
The advantage of quantum mechanics over classical theories for certain information processing tasks lies precisely in the non-locality of quantum correlations.
On the contrary, local realistic (LR) or hidden variable (LHV) theories are described by local objective properties that are independent of observation. LR assumption has experimental consequences providing constraints on the statistics of two or more physically separated systems through Bell inequalities~\cite{PhysicsPhysiqueFizika.1.195}. These inequalities can be violated just by predictions of quantum mechanics.
The structure of non-local correlations is much richer (but also less understood) for multipartite systems than for bipartite ones. In particular, there exist different notions of non-locality by extending the bipartite definition, see for instance~\cite{Brunner:2013est}.
Bell inequality tests to these three-body Higgs decays are also explored in this work.

Since the Higgs boson discovery in 2012, the determination of its properties is part of the major experimental program of ATLAS and CMS Collaborations. 
In the Standard Model (SM) context, the $H\to\gamma\dilep$ decays were discussed and related to two-body $H\to\gamma\gamma,\gamma Z$ decays in~\cite{Chen:2012ju,Dicus:2013ycd,Passarino:2013nka,Kachanovich:2021pvx}.
These decays were also examined within various beyond SM theories~\cite{Corbett:2021iob,VanOn:2021myp,Phan:2022hee,Hue:2023tdz,Tran:2023vgk}.
There are proposals to study CP properties of the Higgs boson in these three-body decays via the forward-backward asymmetry~\cite{Korchin:2014kha,Kachanovich:2020xyg,Aakvaag:2023xhy} and other polarization-dependent observables were introduced in~\cite{Akbar:2014pta,Ahmed:2023vyl}. 
Also, the Higgs radiative decay to a fermion pair plus a photon was analized in~\cite{Han:2017yhy} as a complementary way to measure certain Yukawa couplings and the potential observability of these channels at HL-LHC is evaluated in this reference.
This work analyzes novel observables for these three-body decays and extends the understanding of quantum interactions within such systems.
Concerning the experimental searches, the 13 TeV data analyses for the decay of a Higgs boson in the $\gamma\dilep$ channel were performed by CMS~\cite{CMS:2018myz} and ATLAS~\cite{ATLAS:2021wwb}.

Despite the fact that we do not delve into experimental aspects, this work provides a new conceptual analysis of entanglement and non-locality properties of multipartite states produced in loop-induced Higgs decays.
Predictions of related quantities from Monte-Carlo simulations in complete collider events lie beyond the scope of this work.
Consequently, the aim of this paper is to identify kinematical regions of the phase-space where interesting quantum mechanical measurements might be performed.
To our knowledge, this represents the first analysis of tripartite entanglement in a full 1-loop SM computation.
It is worth noting that the application of this analysis at the detector level would require polarization measurements of the final high-energetic photons (see for instance a related discussion in~\cite{Fabbrichesi:2022ovb}). Although this kind of measurement is not currently available in ATLAS and CMS detectors, in contrast to the case of massive gauge bosons, the LHCb Collaboration performed analysis for photon polarization in $b$-baryon decays~\cite{LHCb:2021byf}. There are also proposals to study CP properties of the Higgs boson through the di-photon decay~\cite{Bishara:2013vya,Gritsan:2022php}.
\newline

The paper is structured as follows: 
\secref{section-observable} provides a comprehensive overview of the 1-loop computation of the helicity amplitudes corresponding to $H\to\gamma\dilep$ decays.
In \secref{section-methods}, we introduce the tripartite density matrix formalism along with the measures associated with entanglement and Bell non-locality. Analytical results can be found in this section.
The \secref{section-reldecays} is devoted to a comparative analysis of entanglement properties of these three-body decays with the related two-body $H\to\dilep,\gamma\gamma,\gamma Z$ decays.
The numerical results of this work are collected in \secref{section-numresults} for each tau-lepton, muon and electron cases. We then summarize the main findings and outline future perspectives in \secref{section-conclus}.
Appendices contain detailed information for the helicity amplitude computation and show the main numerical results using Dalitz plots representation.

\section{Three-body Higgs boson decays}
\label{section-observable}

The present computation is performed in the context of the SM with additional CP-violating terms in the lepton Yukawa sector. Concretely, we consider a generic interaction between the Higgs boson to each lepton $l=e,\mu,\tau$ as
\be
\mL_{H\dilep} = -\frac{\ylep}{\sqrt{2}}\kCP H \bar{\psi_l}(\cCP+i\gamma^5\sCP)\psi_l \,,
\label{HllCPint}
\ee
where $\ylep=\mlep/\vev$ is the Yukawa coupling, with $\mlep$ as the lepton mass and the vev of the Higgs field $\vev=246$ GeV. The magnitude of this interaction and the CP-phase are parametrized by $\kCP\in\Re^+$ and $\dCP\in[0,2\pi]$. The SM is recovered for $\kCP=1$ and $\dCP=0$.

We are interested in quantum entanglement properties of the spin degrees of freedom in the  rare $H\to\gamma\dilep$ decays, then we computed the corresponding helicity amplitudes. We denote the four-momenta of the photon, lepton and antilepton by $k$, $p_-$ and $p_+$ and their helicities, along the direction of motion, as $s_1$, $s_2$ and $s_3$. In addition, $\varepsilon_{s_1}=\varepsilon_{s_1}(k)$ is the polarization vector of the photon and $u_{s_2}=u_{s_2}(p_-)$, $v_{s_3}=v_{s_3}(p_+)$ are the lepton and antilepton spinors (conventions collected in \appref{App-kinem}).

\begin{figure}[h!]
\centering
\begin{tabular}{cccc}
& \includegraphics[width=0.225\textwidth]{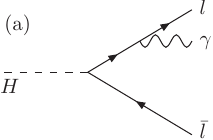} & \includegraphics[width=0.225\textwidth]{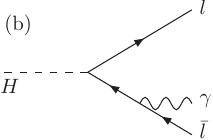} & \\
\addlinespace[4mm]
\includegraphics[width=0.225\textwidth]{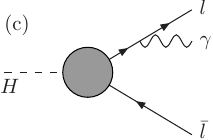} & \includegraphics[width=0.225\textwidth]{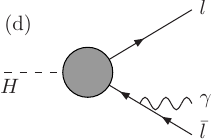} & \includegraphics[width=0.225\textwidth]{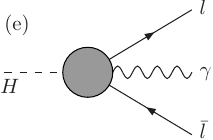} & \includegraphics[width=0.225\textwidth]{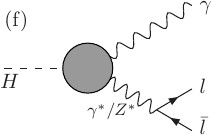} \\
\end{tabular}
\caption{Generic diagrams corresponding to $H\to\gamma\dilep$ decays at $\mO(\hbar)$ in perturbation theory. The gray blobs denote the one-particle-irreducible (1PI) Green functions renormalized in the on-shell scheme.}
\label{diags}
\end{figure}

The generic diagrams corresponding to $H\to\gamma\dilep$ are depicted in \figref{diags}.
Diagrams (a) and (b), in the first row of this figure, correspond to the photon emission process at $\mO(\hbar^0)$, i.e. leading order (LO) in perturbation theory.
This tree level contribution is suppressed by one power of the Yukawa coupling and the resulting helicity amplitude is
\bear
\mM^{\rm Tree}_{s_1s_2s_3} &=& \Ctree\kCP \bar{u}_{s_2}\left( \frac{(\slashed{\varepsilon}_{s_1}^*\slashed{k}+2\varepsilon_{s_1}^*\cdot p_-)(\cCP+i\gamma^5\sCP)}{2k\cdot p_-} \right. \nn\\
&&\left.\hspace{24mm} -\frac{(\cCP+i\gamma^5\sCP)(\slashed{k}\slashed{\varepsilon}_{s_1}^*+2\varepsilon_{s_1}^*\cdot p_+)}{2k\cdot p_+} \right)v_{s_3}  \,,
\label{amptree}
\eear
where the global factor $\Ctree$ is $e\mlep/\vev$ ($e$ as the electromagnetic coupling constant).

On the other hand, diagrams corresponding to the electroweak 1-loop $\mO(\hbar)$ contribution, i.e next-to-leading order (NLO), are schematically represented in the second row of \figref{diags}. The gray blobs denote the one-particle-irreducible (1PI) Green functions renormalized in the on-shell scheme. 
Diagrams (c) and (d) correspond to photon emission, diagram (e) represents the 4-legs 1PI (usually called boxes), and (f) has $H\to\gamma V^*$ as two-body intermediate Higgs boson decay (with $V=\gamma,Z$).
In this $\mO(\hbar)$ computation, the usual linear covariant $R_\xi$-gauge is implemented for the bosonic loops. For the fermionic loops, only the top-quark Yukawa coupling is considered and the rest are neglected. 
In particular, $\mO(\hbar)$ lepton mass effects are relevant just close to the lepton-pair production threshold and we avoid this region considering dilepton invariant mass above $0.1\mh$~\cite{Passarino:2013nka,Kachanovich:2020xyg,Kachanovich:2021pvx}. Hence, there is not suppression with $\ylep$ nor CP-effects in this contribution (in contrast to the tree level).
Also, we have vanishing renormalized 1PI corresponding to 2-legs Green functions $\sum_{H\gamma}$, $\sum_{HZ}$, $\sum_{H\phi_Z}$ and $\sum_{\gamma Z}$, and 3-legs Green function $\Gamma_{H\gamma\phi_Z}$ in this setup\footnote{Concretely, the total 1-loop contributions of the mixing between the Higgs boson with $\gamma$, $Z$ and neutral Goldstone boson $\phi_Z$ are vanishing, the on-shell scheme sets to zero the mixing $\gamma-Z$ at vanishing external momentum, and diagrams type (f) with intermediate $\phi_Z$ are neglected since they are proportional to the lepton mass.}.
Hence the resulting helicity amplitude is written as

\bear
\mM^{\rm 1-loop}_{s_1s_2s_3} &=& \bar{u}_{s_2}\left( (\varepsilon_{s_1}^*\cdot p_-\,\slashed{k}-k\cdot p_-\,\slashed{\varepsilon}_{s_1}^*)(\aone P_R +\bone P_L) \right. \nn\\
&&\left. \hspace{10mm}+(\varepsilon_{s_1}^*\cdot p_+\,\slashed{k}-k\cdot p_+\,\slashed{\varepsilon}_{s_1}^*)(\atwo P_R +\btwo P_L) \right)v_{s_3}  \,,
\label{amploop}
\eear
where $P_{L,R}=(\mathbb{1}\mp\gamma^5)/2$ and the form factors $\aboth$ and $\bboth$ are functions of the momenta $k$, $p_-$ and $p_+$. 
Notice that electroweak (EW) 1-loop corrections involve the Higgs coupling to bosons and to the top-quark, then yield a nonzero decay amplitudes even for vanishing lepton Yukawa.

The full computation using the $R_\xi$-gauge of all diagrams represented in \figref{diags} in the setup described previously, was performed in~\cite{Kachanovich:2020xyg} (see also~\cite{Han:2017yhy}). In particular, the form factors were provided in the corresponding ancillary file, which is implemented for this work.

The diagrams (f) in the second row of \figref{diags} correspond to the two-body intermediate Higgs decay $H\to\gamma V^*\to\gamma\dilep$.
In that case, we have $\aone=\atwo$ and $\bone=\btwo$ since lepton momenta are combined in order to get the intermediate gauge boson momentum $q=p_-+p_+$. The results for the renormalized 3-legs Green function $\Gamma_{H\gamma V^*}$ in the $R_\xi$-gauge can be found in~\cite{Bergstrom:1985hp,Marciano:2011gm,Herrero:2020dtv}.
Notice that the predictions corresponding to $H\to\gamma V^*\to\gamma\dilep$ are very similar for the three flavors because lepton masses are only present in the spinors (leading to negligible effects far to the threshold), there is no tree level $H\dilep$ interaction, lepton Yukawas in the loops are neglected and we have universality in $V\dilep$ interaction.
Particular attention is devoted to the resonant production of the $Z$ boson, for which a non-vanishing decay width $\wz$ is considered by means of a Breit-Wigner distribution in the subprocesses type (f)~\cite{Kachanovich:2020xyg,Kachanovich:2021pvx}.
Other resonant productions decaying into a lepton-pair correspond to the quarkonium states $J/\psi$ and $\Upsilon(nS)$, but they are also rejected by imposing a lower bound ($0.1\mh$) to the dilepton invariant mass~\cite{CMS:2018myz,ATLAS:2021wwb}.

Furthermore, it is interesting to consider the hybrid computation~\cite{Korchin:2014kha,Aakvaag:2023xhy} in which the tree level and the two-body intermediate Higgs decay are combined, i.e. diagrams (a), (b) and (f) of \figref{diags}.
In summary, the three considered computations in this work are:
\bear
\mM_{\rm full} &=& \mM_{\rm (a)}+\mM_{\rm (b)}+\mM_{\rm (c)}+\mM_{\rm (d)}+\mM_{\rm (e)}+\mM_{\rm (f)} = \mM_{\rm Tree}+\mM_{\rm 1-loop} \,,  \nn\\
\mM_{\rm hybrid} &=& \mM_{\rm (a)}+\mM_{\rm (b)}+\mM_{\rm (f)} = \mM_{\rm Tree}+\mM_{\rm (f)} \,,  \nn\\
\mM_{\rm two-body} &=& \mM_{\rm (f)} \,.
\label{kindcomputations}
\eear

Helicity amplitudes in \eqrefs{amptree}{amploop} are Lorentz invariant and they can be written in terms of the Mandelstam variables $s=(p_-+p_+)^2$, $t=(k+p_-)^2$ and $u=(k+p_+)^2$, which satisfy the relation $s+t+u=\mh^2+2\mlep^2$.
However, a convenient choice of the reference frame simplifies the resulting expressions. For the present computation, we consider the rest frame of the lepton-pair where the $z$-axis is along the direction of the lepton, the $y$-axis is perpendicular to the decay plane and the photon momentum has positive $x$-component. Then the two relevant kinematical variables are the dilepton invariant mass ($m_{\dilep}=\sqrt{s}$) and the polar angle between photon and lepton ($\thlep$). Details of the kinematics are gathered in \appref{App-kinem}.

The differential decay width for $H\to\gamma\dilep$ is computed as
\be
\frac{d\Gamma_{H\to\gamma\dilep}}{dm_{\dilep}\,d\clep} = \frac{\sqrt{s-4\mlep^2}(\mh^2-s)}{256\pi^3\mh^3}\sum_{s_1,s_2,s_3}\vert\mM_{s_1s_2s_3}\vert^2 \,.
\ee
The last sum involves the 8 helicity amplitudes (see \appref{App-helamps}). The polar angle $\thlep$ belongs to $[0,\pi]$ and the Mandelstam variable $s$ has relevant limits
\be
(0.1\mh)^2 \,\leq\, s=m_{\dilep}^2 \,\leq\, s_{\rm cut}=\mh^2-2\mh E^\gamma_{cut} \,,
\label{mdileprange}
\ee
where a lower cut $E^\gamma_{cut}$ to the photon energy is imposed in order to avoid infrared (IR) divergences~\cite{Kachanovich:2020xyg}. 
We apply $E^\gamma_{cut}=1$ GeV, in the Higgs boson rest frame, for all numerical evaluations in this work and no additional cuts over the final particles are applied. In addition, we consider the following input SM parameters
\bear
&&G_F=\frac{1}{\sqrt{2}\vev^2}=1.16637\cdot10^{-5}\,\text{GeV}^{-2}\,,\quad \mh=125.1\,\text{GeV}\,,  \nn\\
&&\mw=80.385\,\text{GeV}\,,\quad \mz=91.1876\,\text{GeV}\,,\quad \wz=2.495\,\text{GeV}\,,\quad \mtop=172.5\,\text{GeV}\,, \nn\\
&&\mtau=1.776\,\text{GeV}\,,\quad \mmu=0.105\,\text{GeV}\,,\quad \mel=0.511\cdot10^{-3}\,\text{GeV}\,.
\label{SMinputs}
\eear

\begin{figure}[h!]
\centering
\hspace{30mm}\includegraphics[width=0.72\textwidth]{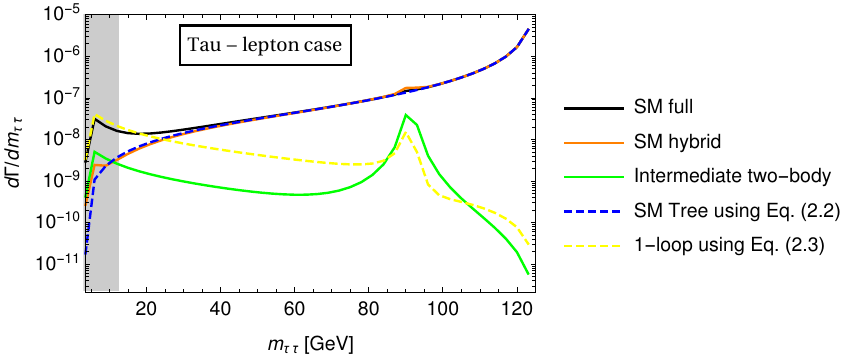}
\newline
\begin{tabular}{ccc}
\addlinespace[2mm]
\includegraphics[width=0.46\textwidth]{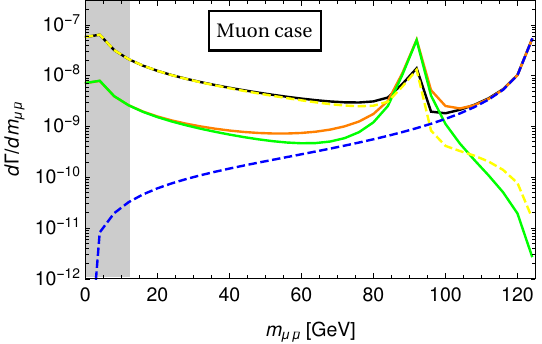} & & \includegraphics[width=0.46\textwidth]{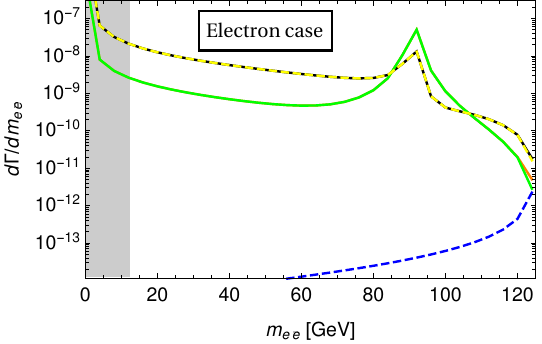}  \\
\end{tabular}
\caption{Differential decay width respect to the dilepton invariant mass for $l=\tau,\mu,e$ in the three kind of computations (solid lines) of \eqref{kindcomputations}. Dashed lines correspond to the tree and complete 1-loop computations in \eqrefs{amptree}{amploop}. SM parameters values $\kCP=1$ and $\dCP=0$ are assumed. Shaded region correspond to $2\mlep\leq m_{\dilep}\leq0.1\mh$.}
\label{anatomyplots}
\end{figure}

The resulting differential decay width respect to the dilepton invariant mass for the three flavors are shown in \figref{anatomyplots}. It is assumed that $\kCP=1$ and $\dCP=0$, i.e. the SM values. Solid lines correspond to the computations in \eqref{kindcomputations}: full (black), hybrid (orange) and two-body intermediate decay (green). Blue and yellow dashed lines account for the tree level in \eqref{amptree} and complete 1-loop (diagrams (c)-(f)) in \eqref{amploop}, respectively. In addition to the relevant range of \eqref{mdileprange}, the range $2\mlep\leq m_{\dilep}\leq0.1\mh$ is included in shaded region for illustrative purposes (just EW 1-loop corrections are considered and we are excluding the contributions associated to $J/\psi$ and $\Upsilon(nS)$ resonances).
See also \tabref{tab:widths} in \appref{App-helamps} for numerical estimations within this setup.

A general conclusion from \figref{anatomyplots} of the full computation is that the interference between tree level and 1-loop contributions is negligible, or in other words, either dominates tree level or dominates 1-loop. 
The only exception is the tau-lepton case (upper plot) near the $Z$-pole peak, where the boson resonance of the 1-loop contribution interferes constructively with the tree-level. In particular, the tree level contribution strongly dominates for energies above 30 GeV in this tau-lepton case.
Regarding the muon case (left-lower plot), the 1-loop controls the behavior up to 100 GeV and the tree level does in the high dimuon invariant mass.
The electron case (right-lower plot) is clearly dominated by the 1-loop contribution in the whole range.
By construction in \eqref{kindcomputations}, the hybrid computation (orange solid lines) are very close to the blue dashed lines when the tree level dominates but they are close to the two-body intermediate (green solid) when the tree level is negligible.

\section{3-qubit formalism}
\label{section-methods}

We are interested in quantum entanglement properties of the spin degrees of freedom corresponding to the final particles in the process $H\to\gamma\dilep$.
Since photons have two transverse polarizations and leptons have two helicities along the momentum direction, they correspond to qubits in the Quantum Information language.
This 3-qubit system is described by the pure state $\vert\psi\rangle$, which is expanded using the helicity amplitude basis $\{+,-\}\otimes\{+,-\}\otimes\{+,-\}$. 
Therefore, the associated 8$\times$8 density matrix $\rho=\vert\psi\rangle\langle\psi\vert$ for a 3-qubit system, in terms of the helicity amplitudes of the process, is
\be
\langle s_1\,s_2\,s_3\vert \rho \vert \tilde{s}_1\,\tilde{s}_2\,\tilde{s}_3\rangle = \left(\sum_{s_1,s_2,s_3}\vert\mM_{s_1s_2s_3}\vert^2\right)^{-1} \mM_{s_1s_2s_3}\mM^\dagger_{\tilde{s}_1\tilde{s}_2\tilde{s}_3}
\label{rhospin}
\ee
where the first factor in the r.h.s is the total unpolarized square amplitude and determines the normalization Tr[$\rho$]=1.

An important question is to recognize and quantify the entanglement in a given quantum state. Concurrence is one of the well-defined quantitative measures for entanglement/separability criteria~\cite{Hill:1997pfa,Horodecki:2009zz}.
However multipartite systems have richer structure than bipartite ones. In particular, just the so-called \textit{genuine} separability is a direct generalization of the bipartite separability and there are many types of \textit{partial} separability.
Following~\cite{Sakurai:2023nsc,Bernal:2023xzp} to this 3-qubits system, just one bipartition (among 1-2 or 2-1) is relevant since $\mC_{i|jk}=\mC_{jk|i}$. Then we only consider the one-to-other concurrences
\be
\mC_{jk|i} = \mC_{i(jk)} = \sqrt{2(1-\Tr[\rho_{jk}^2])} \,,
\label{eq:one-to-other}
\ee
where $\rho_{jk}$ is the reduced density matrix of subsystem $jk$ by tracing over particle $i$, i.e. $\rho_{jk}=\Tr_i[\rho]$.
The relevance of this quantifier is that a state described by $\rho$ is biseparable if and only if $\mC_{jk|i}=0$.

The one-to-other concurrences define the concurrence triangle and the corresponding area represents a measure of the genuine entanglement of this 3-qubit system (GTE) by computing
\be
\mF_3 = \sqrt{\frac{16}{3}S(S-\mC_{23|1})(S-\mC_{31|2})(S-\mC_{12|3})} \,,
\label{eq:aGTE}
\ee
where $S=(\mC_{23|1}+\mC_{31|2}+\mC_{12|3})/2$ is the semiperimeter of the concurrence triangle~\cite{Jin:2022kxb}.
Genuine multipartite entanglement arises in a quantum state when it cannot be expressed as a convex combination of biseparable states, or equivalently, the system is entangled respect to all bipartitions of the parties.

In addition, the entanglement between two individual particles is evaluated by the one-to-one concurrences $\mC_{jk}$, which are obtained from the eigenvalues $\eta^{jk}$'s (in decreasing order) of the matrix $R_{jk}=\sqrt{\sqrt{\rho_{jk}}(\sigma_2\otimes\sigma_2)\rho_{jk}^*(\sigma_2\otimes\sigma_2)\sqrt{\rho_{jk}}}$, as follows
\be
\mC_{jk} = {\rm Max}\{0,\eta_1^{jk}-\eta_2^{jk}-\eta_3^{jk}-\eta_4^{jk}\} \,.
\label{eq:one-to-one}
\ee

Next, we can analytically compute the previous quantifiers for the final state of the $H\to\gamma\dilep$ decay. The 8 helicity amplitudes are written in terms of the generic 1-loop form factors and are collected in \eqref{helamps}.
Remember that we impose a lower limit in \eqref{mdileprange} to the dilepton invariant mass, then $\mlep\ll\sqrt{s}$ regime is satisfied for electron and muon cases, whereas tiny tau-lepton mass effects are present near this lower limit.
Hence, in the high dilepton invariant mass regime respect to $\mlep$, we can neglect lepton masses (except in the $\Ctree$ factor) and a compact expression combining \eqrefs{amptree}{amploop} can be written for the pure final state:

\bear
\hspace{-3mm}\vert\psi\rangle &\simeq& \frac{-i}{N}\left( 8\Ctree\kCP e^{-i\dCP}\frac{s}{(\mh^2-s)\slep}\vert+++\rangle +\atwo(\mh^2-s)\sqrt{s}(1+\clep)\vert++-\rangle \right.  \nn\\
&&\left.\hspace{5mm} +\bone(\mh^2-s)\sqrt{s}(1-\clep)\vert+-+\rangle +8\Ctree\kCP e^{i\dCP}\frac{\mh^2}{(\mh^2-s)\slep}\vert+--\rangle \right.  \nn\\
&&\left.\hspace{5mm} +8\Ctree\kCP e^{-i\dCP}\frac{\mh^2}{(\mh^2-s)\slep}\vert-++\rangle -\aone(\mh^2-s)\sqrt{s}(1-\clep)\vert-+-\rangle \right.  \nn\\
&&\left.\hspace{5mm} -\btwo(\mh^2-s)\sqrt{s}(1+\clep)\vert--+\rangle +8\Ctree\kCP e^{i\dCP}\frac{s}{(\mh^2-s)\slep}\vert---\rangle \right)  
\label{psimlzero}
\eear
where the normalization factor is
\bear
N^2 &=& 128\Ctree^2(\kCP)^2\frac{\mh^4+s^2}{(\mh^2-s)^2\slep^2}  \nn\\
&&+ \left((|\aone|^2+|\bone|^2)(1-\clep)^2+(|\atwo|^2+|\btwo|^2)(1+\clep)^2\right)(\mh^2-s)^2s \,.
\eear

In the vanishing lepton mass limit, the tree level contribution has an IR divergence when the photon is collinear to lepton or antilepton ($\thlep=0,\pi$), and the upper limit in \eqref{mdileprange} must be also imposed. Keeping lepton mass, the collinear configuration is well-defined, as shown in \eqref{ampscol}.
Notice that each helicity amplitude receive contributions either from tree level or 1-loop in the massless lepton regime. In particular, interference terms appear proportional to lepton masses and can be considered negligible. 

Furthermore, the previous quantifiers in \eqrefs{eq:one-to-other}{eq:one-to-one} can be computed analytically in some particular cases.
Considering just the tree level, i.e. \eqref{amptree}, a direct computation yields to 
\bear
&&\CltoAlbar^{\rm Tree} = \ClbartoAl^{\rm Tree} = 1\,,  \nn\\
&&\CAtollbar^{\rm Tree} = \frac{\left(\mh^2-s\right) \left(s-\costhlep^2 \left(s-4 \mlep^2\right)\right)^{1/2}}{\costhlep^2 \left(4 \mlep^2-s\right) \left(s \left(s-8 \mlep^2\cosCP^2\right)+\mh^4\right)+s \left(8 \mlep^2\cosCP^2 \left(4 \mlep^2- s\right)+\mh^4-8 \mh^2 \mlep^2+s^2\right)}\times  \nn\\
&& \hspace{12mm}\times\left(\costhlep^2 \left(4 \mlep^2-s\right) \left(s \left(s-16 \mlep^2\cosCP^2\right)+\mh^4+2 \mh^2 s\right) \right.  \nn\\
&&\left. \hspace{17mm}+s \left(16 \mlep^2\cosCP^2 \left(4 \mlep^2-s\right)+\mh^4+2 \mh^2 \left(s-8 \mlep^2\right)+s^2\right)\right)^{1/2}  \,,
\label{eq-Conetoother}
\eear
with $\costhlep=\clep$, $\cosCP=\cCP$ and we keep $\mlep$ terms.

Notably, the tree level concurrences of the lepton and antilepton respect to the other particles attain the maximal theoretical value 1 independently on the kinematics.
In that case, the $\aGTE$ measure of \eqref{eq:aGTE} is very similar to $\CAtollbar$ since they have values in the interval $[0,1]$.

On the other hand, the photon-to-dilepton concurrence dependence on the CP-phase $\dCP$ is suppressed by the lepton mass, and then negligible in the relevant energy range of \eqref{mdileprange}. Also, the global factors $\Ctree$ and $\kCP$ are cancel out due to the normalization of the state $\rho$\footnote{In general, global factors in the helicity amplitudes disappear after impose the condition $\Tr[\rho]=1$. Including the 1-loop form factors, $\Ctree$ and $\kCP$ are present in the concurrences.}.
In particular, $\CAtollbar^{\rm Tree}$ never vanishes but in the high energy regime it simplifies to
\be
\CAtollbar^{\rm Tree}\vert_{\mlep\ll\sqrt{s}} = \frac{\mh^4-s^2}{\mh^4+s^2} \,,
\label{eq-CAtollbar}
\ee
which is very close to zero in the cut energy $\sqrt{s}_{\rm cut}$ and we almost have the biseparable state\\ $\sim(\vert+\rangle+\vert-\rangle)\otimes(\vert++\rangle+e^{2i\dCP}\vert--\rangle)$, where the normalization factor is omitted. 
On the contrary, if the photon is collinear with lepton or antilepton, $\CAtollbar^{\rm Tree}$ reaches the maximal value 1. 

Now considering just the 1-loop contribution and neglecting lepton mass terms, the one-to-other concurrences are
\bear
&&\CltoAlbar^{\rm 1-loop} = \ClbartoAl^{\rm 1-loop} = \frac{2\left((|\aone|^2(1-\costhlep)^2+|\atwo|^2(1+\costhlep)^2)(|\bone|^2(1-\costhlep)^2+|\btwo|^2(1+\costhlep)^2)\right)^{1/2}}{(|\aone|^2+|\bone|^2)(1-\costhlep)^2+(|\atwo|^2+|\btwo|^2)(1+\costhlep)^2} \,,  \nn\\
&&\CAtollbar^{\rm 1-loop} = \frac{2\left\vert \aone\bone(1-\costhlep)^2-\atwo\btwo(1+\costhlep)^2\right\vert}{(|\aone|^2+|\bone|^2)(1-\costhlep)^2+(|\atwo|^2+|\btwo|^2)(1+\costhlep)^2}  \,,
\label{ConetootherLoop}
\eear
We found that the 1-loop concurrences of the lepton and antilepton respect to the other particles never vanish and achieve the maximal value 1 if $\aone=\atwo$ and $\bone=\btwo$ (as in the two-body intermediate decay).
In that case, the photon-to-dilepton concurrence vanishes when $\costhlep=0$ and we have the biseparable state $\sim(\vert+\rangle-\vert-\rangle)\otimes(\aone\vert+-\rangle+\bone\vert-+\rangle)$ where the normalization factor is omitted. 

In general, considering both tree level and 1-loop contributions, the conditions for genuine entanglement using the concurrence vector formalism to this 3-qubit system~\cite{Bernal:2023xzp} are $\{q_0,q_1,q_2\}=\{0,0,0\}$, where
\bear
q_0 &=& \frac{1}{N^2}\Ctree\kCP e^{-i\dCP}\sqrt{\frac{s}{1-\costhlep^2}} \left(\bone(1-\costhlep)\mh^2+\btwo(1+\costhlep)s\right) \,,\nn\\
q_1 &=& -\frac{1}{N^2}\Ctree\kCP e^{i\dCP}\sqrt{\frac{s}{1-\costhlep^2}} \left(\aone(1-\costhlep)\mh^2+\atwo(1+\costhlep)s\right) \,,\nn\\
q_2 &=& \frac{1}{8N^2}\left(64\Ctree^2(\kCP)^2\frac{\mh^2+s}{(\mh^2-s)(1-\costhlep^2)}\right. \nn\\
&&\hspace{12mm}\left.-(\mh^2-s)^2s\left(\aone\bone(1-\costhlep)^2-\atwo\btwo(1+\costhlep)^2\right)\right) \,,
\eear
for which we neglected lepton mass terms and used \eqref{psimlzero}.
Of course, our previous findings in \eqrefs{eq-Conetoother}{ConetootherLoop} for either tree level or 1-loop contribution are recovered from these conditions. 
The $q_0=0$ and $q_1=0$ equations relate the form factors to each other, and $q_2=0$ establish a relation between them and the tree level factors. 
For each kind of computation (full or two-body intermediate decay), we have non-trivial dependence of the form factors with $s$ and $\clep$, and very particular kinematical configurations could correspond to biseparable states (the numerical analysis is developed in \secref{section-numresults}).

Concerning the one-to-one concurrences, they require a 4$\times$4 diagonalization and analytical expressions result just by neglecting lepton mass terms. 
In that regime, both photon-to-lepton $\CAl$ and photon-to-antilepton $\CAlbar$ exactly vanish for both tree level and 1-loop contributions (then lepton mass terms in the spinors will be relevant in the numerical analysis) and the lepton-to-antilepton $\Cllbar$ for the tree level is very compact
\bear
&&\CAl^{\rm Tree,1-loop}\vert_{\mlep\ll\sqrt{s}} = \CAlbar^{\rm Tree,1-loop}\vert_{\mlep\ll\sqrt{s}} = 0  \nn\\
&&\Cllbar^{\rm Tree}\vert_{\mlep\ll\sqrt{s}} = \frac{2\mh^2s}{\mh^4+s^2} \,.
\label{eq-Cllbar}
\eear
Observe that $\Cllbar^{\rm Tree}$ never vanishes and is very close to 1 in the cut energy $\sqrt{s}_{\rm cut}$. 
The expression corresponding to the 1-loop contribution is not illuminating and it is omitted, however when the photon is collinear with the lepton or antilepton, it vanishes. 

The previous two classes of entanglement (one-to-other and one-to-one) are inequivalent. In fact, the $i$-to-$jk$ entanglement limits the entanglements $i$-$j$ and $i$-$k$ by means of the Coffman-Kundu-Wootters (CKW) monogamy inequality~\cite{Coffman:1999jd}
\be
0 \leq t_3=\mC_{jk|i}^2-\mC_{ij}^2-\mC_{ik}^2
\label{CKWineq}
\ee
where the rhs corresponds to the three-tangle measure, which is the same for all permutations of subsystem indices. For the present rare decays, \eqrefs{eq-Conetoother}{eq-Cllbar} show that the three-tangle is essentially the same as the square of the $\CAtollbar$ concurrence.

Finally, 
regarding the Bell non-locality, local realism (LR) had experimental consequences that can be tested through the inequality of the expectation value of certain Bell operator $\mB$:
\be
\langle\mB\rangle = \Tr[\rho\cdot\mB] \leq \beta_{\rm LR} \,,
\label{Bineq}
\ee
where $\beta_{\rm LR}$ is the locally realist bound and, if an experiment shows $\langle\mB\rangle>\beta_{\rm LR}$, then there is a non-local communication between different particles of a composed system.
For the 2-qubit case, the Clauser-Horne-Shimony-Holt (CHSH) operator~\cite{CHSH} yields to $\beta_{\rm LR}=2$ and it is the only relevant (optimal) Bell operator for this case.

In the multipartite case, non-locality displays a more complex structure, then much richer, and the characterization of multipartite non-local correlations results a challenging problem.
In particular, there exist different notions of non-locality arising as extensions of the bipartite definition~\cite{Brunner:2013est}.
In this work, we implement the 3-qubit Mermin operator~\cite{Mermin:1990vxe}
\be
\mM_3 = \hat{a}_1\otimes\hat{b}_1\otimes\hat{c}_2 +\hat{a}_1\otimes\hat{b}_2\otimes\hat{c}_1 +\hat{a}_2\otimes\hat{b}_1\otimes\hat{c}_1 -\hat{a}_2\otimes\hat{b}_2\otimes\hat{c}_2 \,,
\label{M3def}
\ee
with $\hat{a}_i=\vec{a}_i\cdot\vec{\sigma}$, $\hat{b}_i=\vec{b}_i\cdot\vec{\sigma}$ and $\hat{c}_i=\vec{c}_i\cdot\vec{\sigma}$ are Hermitian operators acting on the Hilbert spaces of the photon, lepton and antilepton, respectively.
The maximum value of this operator in a local realistic theory is $\langle\mM_3\rangle_{\rm LR}=2$, whereas the quantum maximum value is $\langle\mM_3\rangle_{\rm QM}=4$.
There are other options for the Bell operator in the literature, as for example $\mS_3$ proposed by Svetlichny in the seminal work~\cite{Svetlichny:1987bqj}. 
Notice that Mermin inequality can be violated by biseparable states. This is in contrast to the $\mS_3$ case~\cite{Brunner:2013est,Alsina_2016}, which yields to a stronger inequality but it is not a necessary requirement for genuine tripartite non-locality. In addition, the ratio $\langle\mB\rangle_{\rm QM}/\langle\mB\rangle_{\rm LR}$ is greater in the Mermin case.
We found that both operators essentially have the same behaviour for these three-body Higgs boson decays. We also verified that some final state configurations result in predictions compatible with LR using $\mS_3$ in \eqref{Bineq} but violates this inequality using $\mM_3$. Then we just present the numerical results of the Mermin operator in \secref{section-numresults}.

Observe that the expectation value of the Mermin operator in \eqref{M3def} involves a maximization over the six directions corresponding to $\vec{a}_{1,2}$, $\vec{b}_{1,2}$ and $\vec{c}_{1,2}$. As far as we know, there is no closed analytical way to determine this maximum for an arbitrary 3-qubit system, as the 2-qubit case using the Horodecki condition~\cite{Horodecki:1995nsk} for the CHSH operator.

\section{Related bipartite Higgs boson decays}
\label{section-reldecays}

In this section, we compare the entanglement properties of our process of interest $H\to\gamma\dilep$ respect to related bipartite Higgs boson decays, that were previously studied in the literature. 
On the one hand, the photon emission process, diagrams (a)-(b) in \figref{diags}, can be connected with $H\to\dilep$. 
On the other hand, diagrams type (f) are associated to the $H\to\gamma V$ processes ($V=\gamma,Z$).
We will relate the one-to-one and one-to-other concurrences of the previous section with those of the bipartite $H\to\dilep$ and $H\to\gamma V$ decays, respectively.
Some comments are in order: the dilepton decay channel is a well defined observable and $H\to\gamma\dilep$ is part of its NLO corrections. However, $H\to\gamma V^*\to\gamma\dilep$ is unphysical (in particular, gauge-dependent) and it can be extracted as pseudo-observable by imposing kinematical cuts on the dilepton invariant mass~\cite{Passarino:2013nka,Kachanovich:2020xyg,CMS:2018myz,ATLAS:2021wwb}. 

\subsection{$H\to\dilep$ and CP-effects}
\label{section-CPeffects}

Entanglement properties of $H\to\dilep$ were analyzed for the tau-lepton case in~\cite{Fabbrichesi:2022ovb,Altakach:2022ywa}. The CP-violating interaction, as in \eqref{HllCPint}, was also treated in these references. The authors conclude that the concurrence of the ditau system is maximal regardless of the CP-phase and its  determination was obtained by a direct fit of the entries in the correlation matrix of the 2-qubit final state.
In addition, CP properties of the lepton Yukawa sector were measured for the $H\to\ditau$ decay by CMS~\cite{CMS:2021sdq} and ATLAS~\cite{ATLAS:2022akr} and there are proposal to study them through the forward-backward asymmetry in $H\to\gamma\dilep$ decays~\cite{Korchin:2014kha,Aakvaag:2023xhy}.

Now we focus on the tree level contribution since the 1-loop does not depend on $\kCP$ nor $\dCP$. 
The reduced density matrix $\rho_{\dilep}$ of the dilepton subsystem, after tracing over the photon helicity, can be decomposed as
\be
\rho_{\dilep} = \frac{1}{4}\left( \mathbb{1}\otimes\mathbb{1} +\sum_{i=1}^3A_i\sigma_i\otimes\mathbb{1} +\sum_{j=1}^3B_j\mathbb{1}\otimes\sigma_j +\sum_{i,j=1}^3C_{ij}\sigma_i\otimes\sigma_j \right)
\ee
where the coefficients $A$ and $B$ are the spin polarizations and $C$ is the spin correlation matrix of the resulting bipartite subsystem. 
For these rare decays, neglecting the lepton mass, the $A$ and $B$ coefficients vanish and the correlation matrix results
\be
C=\begin{pmatrix}
\cos(2\dCP)\frac{2\mh^2s}{\mh^4+s^2} & \sin(2\dCP)\frac{2\mh^2s}{\mh^4+s^2} & 0 \\
\sin(2\dCP)\frac{2\mh^2s}{\mh^4+s^2} & -\cos(2\dCP)\frac{2\mh^2s}{\mh^4+s^2} & 0 \\
0 & 0 & 1 \\
\end{pmatrix} \,.
\ee

The concurrence of this dilepton subsystem was presented in \eqref{eq-Cllbar}. The dependence on the CP-phase in the correlation matrix is missing in this entanglement measure, but sensitivity to $\dCP$ could be obtained by a direct fit of the entries in the correlation matrix, as in the case of $H\to\ditau$.

We are analyzing novel observables for these tripartite Higgs boson decays. In the previous section, we found that CP-effects are suppressed by the lepton mass for the one-to-other concurrences in \eqref{eq-Conetoother}. In particular, for the relevant dilepton invariant mass range of \eqref{mdileprange}, they are roughly independent of $\kCP$ and $\dCP$. In other words, this kind of quantifiers are not sensitive to the new physics introduced by these parameters. 
Hence, we assume $\kCP=1$ and $\dCP=0$, i.e. SM computation, for the rest of this manuscript.

\subsection{$H\to\gamma Z$ and post-decay entanglement}
\label{section-postdecay}

Other related bipartite decays are $H\to\gamma\gamma$ and $H\to\gamma Z$, associated to diagrams type (f) of \figref{diags}. 
From the experimental point of view, the $H\to\gamma\gamma$ was one of the golden decay channels for the Higgs boson discovery.
The latest ATLAS and CMS measurements of the Higgs boson mass in this channel are presented in~\cite{CMS:2020xrn,ATLAS:2023owm} and the corresponding cross-section can be found in~\cite{CMS:2022wpo,ATLAS:2023tnc}.
On the other hand, a recent combined analysis of ATLAS and CMS found evidence for $H\to\gamma Z$ decay~\cite{ATLAS:2023yqk}, which agrees with the SM theoretical expectation within 1.9 standard deviations.

Entanglement properties of the diphoton channel was previously studied in~\cite{Fabbrichesi:2022ovb}. This final state is maximally entangled, i.e. the concurrence attains the theoretical maximum 1. Furthermore, the corresponding CHSH operator saturates the Cirelson bound~\cite{Cirelson:1980ry} and the Bell inequality is maximally violated.

On the other hand, requiring dilepton invariant mass close to the $Z$-pole peak, the pseudo-observable $H\to\gamma Z^*$ can be enhanced. This final state corresponds to a bipartite system composed by one qubit and one qutrit. This state was studied in~\cite{Morales:2023gow} when it is coming from vector boson scattering and, as far as we know, this is the first time that entanglement properties of $H\to\gamma Z$ are studied. The SM amplitude of this decay is expressed as
\be
\mM_{H\to\gamma Z} = V_{H\gamma Z}(q^\mu k^\nu-g^{\mu\nu}k\cdot q)\varepsilon_{s_1}(k)_\mu^*\varepsilon_{\bar{s}_1}(q)_\nu^* \,,
\label{ampHtoAZ}
\ee
where $\{k^\mu,s_1\}$ and $\{q^\nu,\bar{s}_1\}$ are the momentum and helicity of the photon and $Z$ boson, $\varepsilon$'s are their polarization vectors and the 3-legs form factor $V_{H\gamma Z}$ accounts for the 1-loop contribution (that only depends on $k$ and $q$ momenta). The precise dependence of this form factor on the momenta is not relevant for the following discussion. 
In the $\{+,-\}\otimes\{+,0,-\}$ basis, the resulting density matrix is 
\be
\rho_{H\to\gamma Z} = \begin{pmatrix}
0 & 0 & 0 & 0 & 0 & 0 \\
0 & 0 & 0 & 0 & 0 & 0 \\
0 & 0 & 1/2 & 1/2 & 0 & 0 \\
0 & 0 & 1/2 & 1/2 & 0 & 0 \\
0 & 0 & 0 & 0 & 0 & 0 \\
0 & 0 & 0 & 0 & 0 & 0 \\
\end{pmatrix} \,,
\label{rhoAZ}
\ee
which is independent of the form factor (since it is a global factor of the helicity amplitudes). The concurrence for a bipartite system with arbitrary dimension $d_1\otimes d_2$ was defined in~\cite{Rungta_2001} and the theoretical maximum is $\sqrt{2(d-1)/d}$ with $d={\rm min}\{d_1,d_2\}$. 
For this bipartite decay, the concurrence achieves the maximum 1 and the $\gamma Z$ state (coming from the Higgs boson) results maximally entangled. However, the 2$\otimes$3 generalized CHSH operator~\cite{Caban:2008qa,Barr:2021zcp,Morales:2023gow} for this decay never exceeds 2.

The $H\to\gamma Z$ and our process of interest $H\to\gamma\dilep$ allow to test the post-decay entanglement~\cite{Aguilar-Saavedra:2023hss} and autodistillation phenomena~\cite{Aguilar-Saavedra:2024fig,Aguilar-Saavedra:2024hwd}.
Concretely, the concurrence $\mC_{\rm initial}$ of an `initial' bipartite state $\{a,\bar{a}\}$ is computed. The particle $\bar{a}$ decays along the process $\bar{a}\to b_1b_2$ and the entangled `initial' state led to the entangled `final' subsystems $\{a,b_1\}$ and $\{a,b_2\}$ with concurrences $\mC_{\rm final\,1}$ and $\mC_{\rm final\,2}$, respectively. A priori, different amount of entanglement is expected and the autodistillation phenomena arises when final concurrences after decay are greater than the initial one.
In our context, we consider the two-body intermediate Higgs decay into the `initial' state $\gamma Z$ and the subsequent post-decay $Z\to\dilep$. Diagrams type (f), with $Z$ boson as mediator, contribute to the 1-loop amplitude in \eqref{amploop} as
\be
\mM_{\rm (f)}\vert_{\rm Z} = \mM_{\rm prod}^\nu \left(\frac{\sum_{\bar{s}_1}\varepsilon_{\bar{s}_1}(q)_\nu^*\varepsilon_{\bar{s}_1}(q)_\lambda}{q^2-\mz^2+i\mz\wz}\right)\mM_{\rm decay}^\lambda \,,
\ee
where the intermediate $Z$ boson propagator was written in terms of the polarization vectors and $q=p_-+p_+$.
Close to the $Z$-pole, the Narrow-Width Approximation (NWA) can be applied, which mainly replaces the denominator's contribution of this propagator to the concurrences by a delta function, such that only $Z$ boson on-shell (OS) effects remain. In that case, the previous helicity amplitude is splitted as
\be
\mM_{\rm (f)}^{s_1s_2s_3}\vert_{\rm Z(OS)} \propto\sum_{\bar{s}_1}\mM_{H\to\gamma Z}^{s_1\bar{s}_1}\mM_{Z\to\dilep}^{\bar{s}_1s_2s_3} \,,
\label{amp-f-NWA}
\ee
where $\mM_{H\to\gamma Z}^{s_1\bar{s}_1}$ was given in \eqref{ampHtoAZ} and the SM polarized decay amplitudes for $Z\to\dilep$ are
\be
\mM_{Z\to\dilep}^{\bar{s}_1s_2s_3} = \bar{u}_{s_2}\slashed{\varepsilon}_{\bar{s}_1}\left(\sw^2 P_R +(-1/2+\sw^2)P_L\right)v_{s_3} \,.
\ee

Following~\cite{Aguilar-Saavedra:2023hss,Aguilar-Saavedra:2024fig,Aguilar-Saavedra:2024hwd}, we consider the subsystems $\gamma l$ and $\gamma\bar{l}$, by tracing over $\bar{l}$ and $l$, and compute the one-to-other concurrences $\ClbartoAl$ and $\CltoAlbar$.
Notice that the `initial' state $\gamma Z$, described by \eqref{rhoAZ}, is maximally entangled (concurrence equals to 1) and the original autodistillation phenomena is not relevant now. However, 
we arrive to the one-to-other concurrences 
\be
\ClbartoAl^{\rm (f)} = \CltoAlbar^{\rm (f)} =\frac{4\sw^2(1-2\sw^2)}{1-4\sw^2+8\sw^4}\approx 0.976 \,,
\label{eq-Cpostdecay}
\ee
which are very close to the corresponding one of the $\gamma Z$ state, and we conclude that the $Z$ boson decaying into a lepton-pair does not significantly decrease the initial concurrence. 
Notoriously, if we impose the `MaxEnt Principle'~\cite{Cervera-Lierta:2017tdt} to the previous one-to-other concurrences, i.e. demand $\ClbartoAl^{\rm (f)} = \CltoAlbar^{\rm (f)} =1$, the Weinberg angle should satisfy $\sw^2=0.25$, which is surprisingly close to the SM value~\cite{Workman:2022ynf}.

\section{Numerical Results}
\label{section-numresults}

In this section, we present the distributions of the entanglement and Bell non-locality measures in the $[m_{\dilep},\clep]$ plane of the phase-space.
We analyze the three lepton families separately since the 1-loop contribution dominates different energy regimes in each case, and also corresponds to separate experimental channels.
The numerical results correspond to the SM, i.e. $\kCP=1$ and $\dCP=0$ in \eqref{helamps}, including lepton mass effects and using the corresponding form factors for each kind of computation in \eqref{kindcomputations}.
Then, in the plots, the range for $\clep$ is $[0,1]$ since the distributions result symmetric under $\cos\to-\cos$, and the dilepton invariant mass covers the range in \eqref{mdileprange}.

\begin{figure}[h!]
    \centering
\begin{tabular}{ccr}
\includegraphics[width=0.28\textwidth]{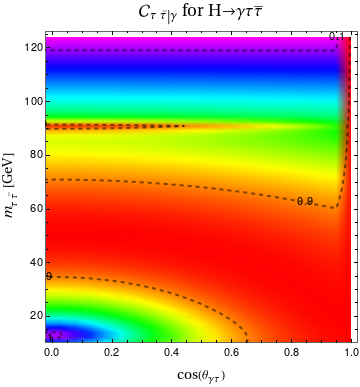} & \includegraphics[width=0.33\textwidth]{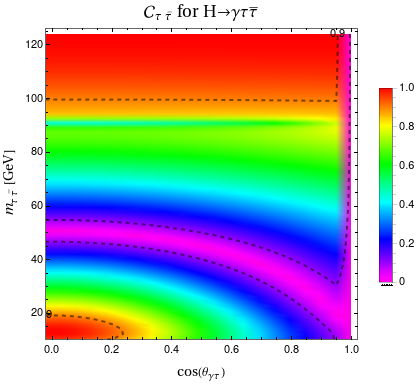} & \includegraphics[width=0.33\textwidth]{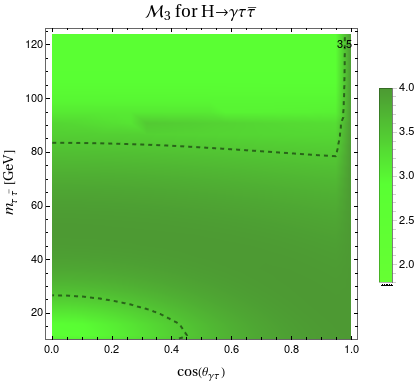} \\
\end{tabular}
\begin{tabular}{cccr}
\includegraphics[width=0.21\textwidth]{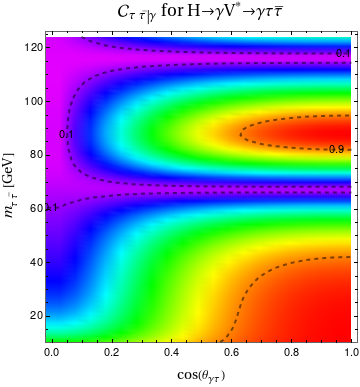} & \includegraphics[width=0.21\textwidth]{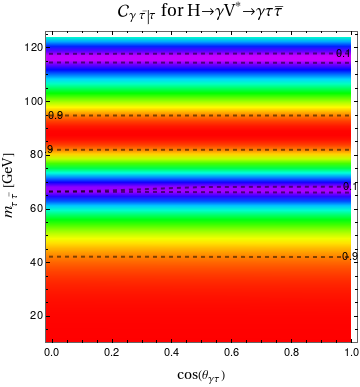} & \includegraphics[width=0.245\textwidth]{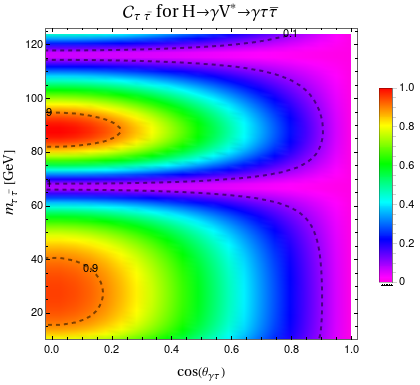} & \includegraphics[width=0.245\textwidth]{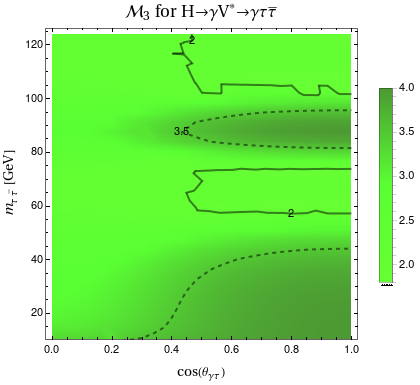} \\
\end{tabular}
\caption{Entanglement and non-locality quantifiers for $H\to\gamma\ditau$ in the $[\mditau,\ctau]$ plane, corresponding to the full computation (first row) and two-body intermediate Higgs decay (second row). Dashed lines are contours corresponding to 0.1 and 0.9 for concurrences and to 2 and 3.5 for Mermin operator.}
\label{tau-plots}
\end{figure}

All the concurrences are defined in such a way that have values in the range $[0,1]$ and the present computation shows that the theoretical minimum and maximum are almost achieved in some particular configurations. 
For the distributions in \figrefs{tau-plots}{el-plots}, the same color-scale is used where the purple(red) regions represent values close to 0(1).
For each point of the phase-space, the CKW inequality of \eqref{CKWineq} was verified. 
Because CP-effects are negligible, the results in this section correspond to the SM computation. Hence $\CltoAlbar$ and $\ClbartoAl$ are the same under transformation $\thlep\to\pi-\thlep$, or equivalently $\clep\to-\clep$. The same occurs for $\CAlbar$ and $\CAl$.
The general behaviour of these quantifiers were treated analytically in \secref{section-methods}. In particular, we have simplified formulas of the concurrences in the limit $\mlep\ll0.1\mh\leq m_{\dilep}$ for both tree level and 1-loop contributions. Of course, the predictions of the 1-loop form factors for each kind of computation in \eqref{kindcomputations} are determined numerically and provide the precise behaviour of these quantifiers in the phase-space.

Concerning the Mermin operator distributions, a green-scale is chosen for the resulting values in the range $[1.8,4]$. In particular, just small regions satisfy the Bell inequality of \eqref{Bineq}, compatible with local-realism, and we conclude that these rare Higgs boson decays are promising observables for test tripartite Bell non-locality in a high-energy regime.

In order to implement the common variables for three-body decays and to be independent of the reference frame, the corresponding distributions using the Dalitz plot representation are collected in \appref{App-Dalitz}.

\subsection{Tau-lepton case}

The $H\to\gamma\ditau$ process is dominated by the 1-loop (tree level) contribution for energies below (above) 30 GeV, as can be seen from the first row of \figref{anatomyplots}. 
Around the $Z$-pole, both contributions interfere constructively but the resonance comes from the 1-loop diagrams type (f).
The resulting one-to-other and one-to-one concurrences $\CAtollbar$ and $\Cllbar$ in the full computation are shown in the first row of \figref{tau-plots}, together with the Mermin operator $\Mertri$ in the $[\mditau,\ctau]$ plane.

The one-to-other concurrence $\CAtollbar$ achieves values close to the theoretical maximum 1 in the red regions, that is roughly for i) energies $\mditau$ in the range $[40,60]$ GeV with $|\ctau|\leq0.5$, ii) energies below 40 GeV with $0.85\leq|\ctau|$, iii) in the directions $\ctau\sim\pm1$.
On the contrary, $\CAtollbar$ is close to zero in the purple region, i.e. for $\mditau\sim\sqrt{s}_{\rm cut}$ and also for $\mditau\sim0.1\mh$ in the central region (where photon and $\tau$ lepton directions are almost orthogonal), reaching minimum values $\sim10^{-2}$.

The one-to-one concurrence $\Cllbar$ have values close 1 for energies above $\sim115$ GeV and also for $|\ctau|\lesssim0.1$ with $\mditau\sim15$ GeV.
This is in contrast to the $\CAtollbar$ for which these regions correspond to the minima.
On the other hand, values of $\Cllbar$ close to zero are located in the purple ring and in the directions $\ctau\sim\pm1$.

As anticipated in \eqref{eq-Conetoother} and \eqref{ConetootherLoop}, the one-to-other concurrences $\CltoAlbar$ and $\ClbartoAl$ are very close to 1 in the whole kinematical plane, where the left- and right-handed form factors are very similar. 
Hence the GTE measure $\aGTE$ has nearly the same distribution as the $\CAtollbar$.
Regarding the one-to-one concurrences $\CAl$ and $\CAlbar$, they are homogeneously distributed around 0 except in the window $0.9\mz\lesssim \mditau\lesssim1.1\mz$ in which take maximal values of $\sim0.3$ at the $Z$-pole, in accordance with \eqref{eq-Cllbar}.

The hybrid computation yields to very similar conclusions respect to the full computation since tree level is dominant for the tau-lepton case.
We found only small variations in the $\mditau\lesssim30$ GeV region since the 1-loop contribution is dominant here.

The second row of \figref{tau-plots} corresponds to the two-body intermediate $H\to\gamma V^*\to\gamma\ditau$ decay.
The resulting quantifiers change respect to the full computation.
Now the one-to-other concurrence $\CAtollbar$ have values close to 1 for $0.6\lesssim|\ctau|$ when $\mditau\lesssim30$ GeV or $0.95\mz\lesssim \mditau\lesssim1.05\mz$. Its minima are located in the central region $|\ctau|\lesssim0.1$ for energies above 70 GeV and also in the bands corresponding to $\mditau\sim70$ and 115 GeV (independently of the angle).

In addition, the one-to-other concurrences $\CltoAlbar$ and $\ClbartoAl$ distributions are not longer homogeneously distributed around 1 and exhibit bands in energy (independent of $\thtau$).
In particular, the maxima corresponds to $\mditau\lesssim30$ GeV and $0.95\mz\lesssim \mditau\lesssim1.05\mz$. The later is expected since the post-decay process takes place as in \eqref{eq-Cpostdecay}.
The minima are achieved for $\mditau\sim66$ and 116 GeV.
Due to these modifications, $\aGTE$ measure also changes but still essentially follows the $\CAtollbar$ distribution with broader purple regions.

The one-to-one concurrence $\Cllbar$ concentrates the maxima in regions with $|\ctau|\lesssim0.2$ and energies near to 30 GeV and $Z$-pole. 
The minima are located $0.9\lesssim|\ctau|$ or $\mditau\sim66$ and 116 GeV.
The one-to-one concurrences $\CAl$ and $\CAlbar$ are homogeneously distributed around 0 even at the $Z$-pole region.

In summary, the full computation exhibit broader regions of high entanglement (concurrences close to 1) respect to the two-body intermediate Higgs decay except near the $Z$-pole with $0.6\lesssim|\ctau|$ for $\CAtotautau$, and $|\ctau|\lesssim0.2$ for $\Ctautau$.

Finally, regarding the Bell non-locality in this decay, the expectation value of the Mermin operator in \eqref{M3def} is shown in the last column of \figref{tau-plots} for the full and two-body intermediate computations.
Remember that for each point of the phase-space, a maximization over six directions is performed.
We found that within the full computation, the minimal value is $\sim2.82$ and then \eqref{Bineq} is violated in the whole plane. In addition, the theoretical maxima 4 is reached when photon is mostly collinear with lepton or antilepton and in the lower-right corner.
For the $H\to\gamma V^*\to\gamma\ditau$ computation, in general the expectation value is decreased, keeping maximal values in the lower-right corner. Observe that there are regions with values lower than 2, i.e. compatible with LR, 
reaching minimum $\sim1.82$.

\begin{figure}[h!]
    \centering
\begin{tabular}{ccr}
\includegraphics[width=0.28\textwidth]{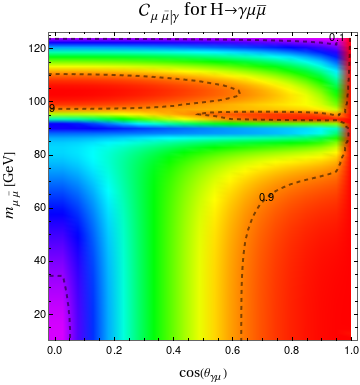} & \includegraphics[width=0.33\textwidth]{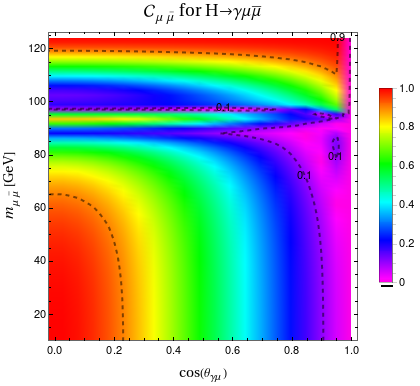} & \includegraphics[width=0.33\textwidth]{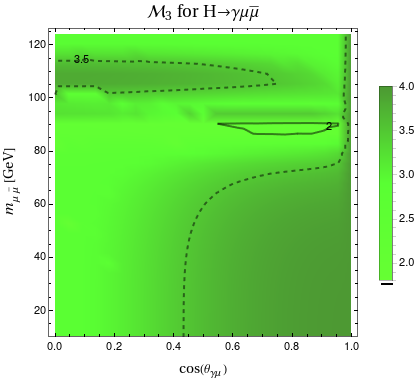} \\
\end{tabular}
\begin{tabular}{cccr}
\includegraphics[width=0.21\textwidth]{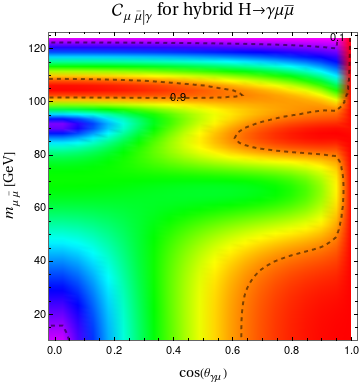} & \includegraphics[width=0.21\textwidth]{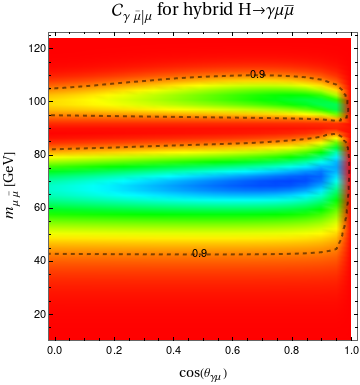} & \includegraphics[width=0.245\textwidth]{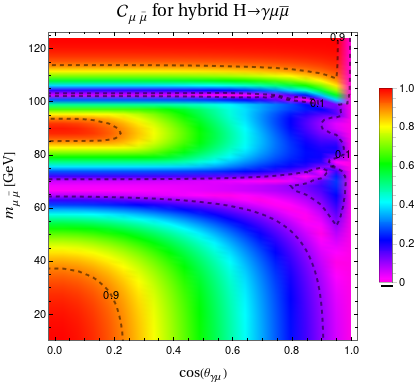} & \includegraphics[width=0.245\textwidth]{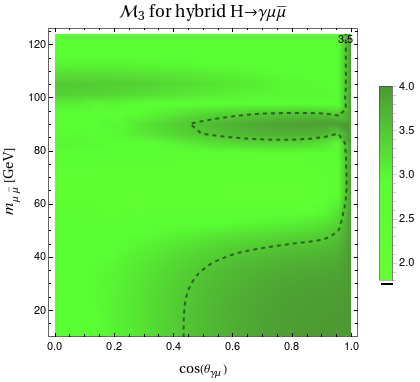} \\
\end{tabular}
\caption{Entanglement and non-locality quantifiers for $H\to\gamma\dimu$ in the $[\mdimu,\cmu]$ plane, corresponding to the full (first row) and hybrid (second row) computations. Dashed lines are contours corresponding to 0.1 and 0.9 for concurrences and to 2 and 3.5 for Mermin operator.}
\label{mu-plots}
\end{figure}

\subsection{Muon case}

For $H\to\gamma\dimu$ process, the tree level contribution is dominant just for energies above 100 GeV.
Hence the resulting distributions are changed respect to the tau-lepton case.

The first row of \figref{mu-plots} collects the $\CAtollbar$, $\Cllbar$ and $\Mertri$ distributions within the full computation.
The one-to-other concurrence $\CAtollbar$ attains maxima (almost 1) for i) energies below 40 GeV and $0.85\lesssim|\cmu|$, and also for ii) energies in the range $[100,110]$ GeV and $|\cmu|\lesssim0.2$.
In particular, the central region with $|\cmu|\lesssim0.2$ and high energy region above 100 GeV are drastically modified respect to the tau-case.
In addition, the one-to-other concurrences $\CltoAlbar$ and $\ClbartoAl$ decreases up to 0.6 in the window $0.9\mz\lesssim\mdimu\lesssim1.1\mz$ but they are uniformly close to 1 in the rest of the plane (as in the tau-case). Even so, the $\aGTE$ measure has distribution almost alike the $\CAtollbar$ one.

The one-to-one concurrence $\Cllbar$ reaches values close to 1 for $\mdimu$ above 120 GeV and for $|\cmu|\lesssim0.05$ with energies below 30 GeV, while values close to zero when $|\cmu|\sim1$. 
The other two one-to-one concurrences are very similar to the $\tau$-case.

The results corresponding to $H\to\gamma V^*\to\gamma\dimu$ are the same as the tau-lepton case (second row of \figref{tau-plots}) since muon mass effects in this two-body intermediate decay are negligible. 
In particular, we have same entanglement distributions as in the second row of \figref{tau-plots} for the considered invariant mass range of \eqref{mdileprange}.

The distributions within the hybrid computation are shown in the second row of \figref{mu-plots} and differ respect to the full computation (first row), in contrast to the $\tau$-case in which both computations are very similar.
In the present case, the hybrid predictions are a sort of transition between the full to the two-body intermediate decay computations. In particular, the red regions are reduced whereas the purple regions increase.

The $\mM_3$ distributions of the last column in \figref{mu-plots} are also changed respect to the tau-lepton case. Now the full computation results in lower values and exhibit a small region satisfying \eqref{Bineq} around the $Z$-pole and $0.6\leq\cmu\leq0.9$ (with minimum $\sim1.9$). Notoriously, the hybrid computation exhibits its largest values in that region. Also, the maxima (almost 4) are located in lower-right corner, i.e. $m_{\dilep}\leq30$ GeV with $0.9\leq|\cmu|$, in both computations.

\begin{figure}[h!]
    \centering
\begin{tabular}{cccr}
\includegraphics[width=0.21\textwidth]{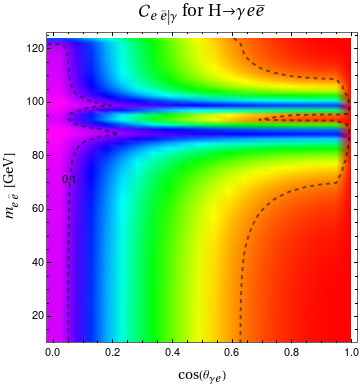} & \includegraphics[width=0.21\textwidth]{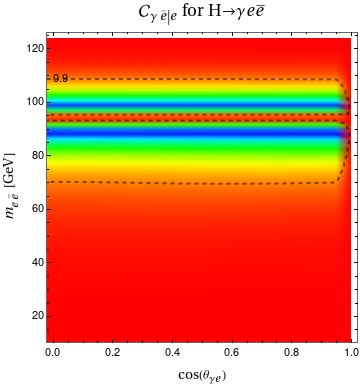} & \includegraphics[width=0.245\textwidth]{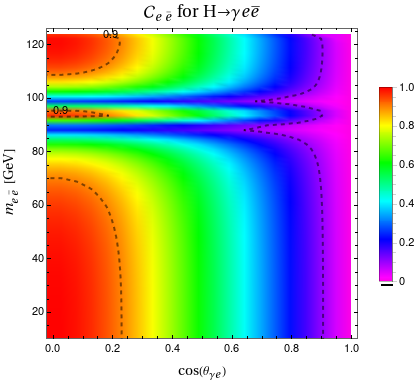} & \includegraphics[width=0.245\textwidth]{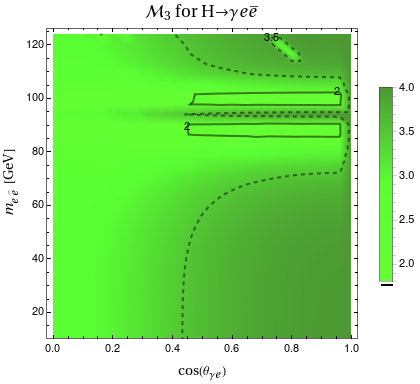} \\
\end{tabular}
\caption{Entanglement and non-locality quantifiers for $H\to\gamma\diel$ in the $[\mdiel,\cel]$ plane corresponding to the full computation. Dashed lines are contours corresponding to 0.1 and 0.9 for concurrences and to 2 and 3.5 for Mermin operator.}
\label{el-plots}
\end{figure}

\subsection{Electron case}

In this channel, the tree-level contribution is negligible since it is suppressed by the electron Yukawa.
Then, the hybrid and two-body intermediate computations are practically identical, leading to distributions as in second row of \figref{tau-plots}.

The full computation distributions are shown in \figref{el-plots}.
The one-to-other concurrence $\CAtollbar$ reaches maxima (almost 1) for $0.8\lesssim|\cel|$ with energies below (above) 40 (115) GeV, while the central region, $|\cel|\lesssim0.1$, corresponds to the minimal values.
The one-to-other concurrences $\CltoAlbar$ and $\ClbartoAl$ are no longer homogeneous around 1, remember \eqref{ConetootherLoop}, and exhibit narrow bands in energy. In particular, values close to 1 are still present at the $Z$-pole (as expected in the post-decay) and for energies below(above) 40(115) GeV.
Therefore, the $\aGTE$ distribution has just tiny variations in the mentioned narrow bands respect to the $\CAtollbar$ distribution.

The one-to-one concurrence $\Cllbar$ attains maxima for $|\cel|\lesssim0.1$ with energies below(above) 40(115) GeV, while the minimal values are achieved for $0.95\lesssim|\cel|$.
The one-to-one concurrences $\CAl$ and $\CAlbar$ are uniformly distributed close to 0 in the whole kinematical plane, as expected from \eqref{eq-Cllbar}.

Finally, the Mermin operator distribution of the last plot in \figref{el-plots} shows values lower than 2 (with minimum $\sim1.68$) for energies $\sim80,100$ GeV and $0.45\leq\cel\leq0.85$. Now, the theoretical maxima 4 is reached in the lower-right and upper-right corners.

\section{Summary and perspectives}
\label{section-conclus}

This work aims to elucidate the mechanisms underlying the generation and distribution of entanglement within the framework of fundamental interactions described by the SM and incorporating CP-violation in the lepton Yukawa sector.
Based on the concurrence and Bell operator definitions for tripartite systems, we explore entanglement and non-locality properties of $H\to\gamma\dilep$ decays (for $l=\tau,\mu,e$). These three-body decays are Yukawa suppressed at leading-order, then electroweak 1-loop corrections are included. They offer a unique opportunity to examine quantum correlations arising at NLO in perturbation theory within the SM.
This paper presents novel observables for these three-body Higgs boson decays and extend our understanding of quantum interactions within such systems.

Our goal was to identify regions of the phase-space where the final particles result entangled after the Higgs boson decay and to determine the feasibility of testing Bell inequality under these kinematical configurations.
By expanding beyond traditional bipartite systems to the three-body final state, we compute various entanglement measures including one-to-other and one-to-one concurrences, the area of the concurrence triangle, the conditions for genuine entanglement of 3-qubit systems using the concurrence vector and the three-tangle measure.
These quantifiers were derived analytically in terms of the generic 1-loop form factors and numerical predictions in the $[m_{\dilep},\clep]$ plane for three kind of computations were also explored. 

We found that the final photon, lepton and antilepton result entangled after the Higgs boson decay, and this also holds by considering the one-to-one and one-to-other subsystems among them. The amount of entanglement depends on the final state kinematical configuration and maximally entangled subsystems appear in certain regions of the phase-space.
Concerning the Bell non-locality, both Mermin and Svetlichny operators for 3-qubit systems were computed. We detect predictions incompatible with local realism in the whole phase-space, except for a few particular configurations, suggesting that $H\to\gamma\dilep$ could serve as an ideal laboratory for testing Bell inequality.
The development of possible experimental implementations is out of the scope of this work  and it is deferred to future study.

Furthermore, we analyzed post-decay entanglement and autodistillation phenomena at dilepton invariant mass close to the $Z$-pole mass. We found that the qubit-qutrit system of gauge bosons in the $H\to\gamma Z$ decay is maximally entangled, with minimal entanglement loss in the 2-qubit subsystems (photon-lepton and photon-antilepton) of the $H\to\gamma Z\to\gamma\dilep$ decay. 
Interestingly, the `MaxEnt Principle' applied to these subsystems favors $\sw^2=0.25$, notoriously close to the SM value.
On the other hand, when introducing CP-violating interactions in the lepton Yukawa sector, we observe that CP-effects on these entanglement measures are suppressed by lepton masses, thus these observables are not suitable for studying such kind of new physics.

Future avenues for exploration include extending this analysis to different three-body decays such as the well-studied $\pi^0\to\gamma e^-e^+$ or hadrons decays into three fermions.
Furthermore, continuing with Higgs boson decays, a natural multipartite extension is to consider the four-fermion channel, constituting a 4-qubit system.

\section*{Acknowledgments}

I thank to Bianca Polari for her encouragement and support throughout this study. 
I am grateful to Alejandro Szynkman for useful comments and suggestions and also to the organizers and participants of the workshop Quantum Observables for Collider Physics, November 2023, funded by the Galileo Galilei Institute for Theoretical physics of the Istituto Nazionale di Fisica Nucleare, for the interesting and inspiring discussions developed there.
The present work has received financial support from CONICET and ANPCyT under project PICT-2021-00374.

\section*{Data Availability Statement}

No Data associated in the manuscript.

\section*{Appendices}
\appendix

\section{Conventions and kinematics}
\label{App-kinem}

The relevant details for the computation of the helicity decay amplitudes are summarized in this appendix. 
Regarding the Dirac algebra, we use the Weyl representation of $\gamma$ matrices. The spinors $u$ and $v$, solutions of the Dirac equation for particle and antiparticle, are also eigenvectors of the helicity operator along the momentum direction $\vec{p}$:
\be
\hat{\Lambda}_{\vec{p}} = \frac{\vec{p}\cdot\vec{S}}{|\vec{p}|} \,,
\ee
where the spin matrices, in terms of the Pauli matrices $\sigma_i$, are
\be
S_x = \begin{pmatrix}
\sigma_1 & 0 \\
0 & \sigma_1 
\end{pmatrix}\,,  \hspace{8mm}  S_y = \begin{pmatrix}
\sigma_2 & 0 \\
0 & \sigma_2  
\end{pmatrix}\,,  \hspace{8mm}  S_z = \begin{pmatrix}
\sigma_3 & 0 \\
0 & \sigma_3 
\end{pmatrix} \,.
\ee

In particular, the spinors have eigenvalues $\pm1/2$ and we follow the Chanowitz convention
\be
\hat{\Lambda}_{\vec{p}}\, u_\lambda(\vec{p}) = \frac{\lambda}{2} u_\lambda(\vec{p}) \quad\text{and}\quad \hat{\Lambda}_{\vec{p}}\, v_\lambda(\vec{p}) = -\frac{\lambda}{2} v_\lambda(\vec{p}) \,.
\ee
For simplicity in the computational basis notation, the subindices of the spinors are twice the spin along the momentum direction. 
The explicit expressions for these spinors can be found in the Appendix A of~\cite{Blasone:2024dud}. Notice that the antiparticle convention implemented here is the opposite to the chosen one in~\cite{Fabbrichesi:2022ovb,Altakach:2022ywa} and this is reflected in a reordering of the density matrix elements when comparing to that references.

\begin{figure}[h!]
\centering
\includegraphics[width=0.6\textwidth]{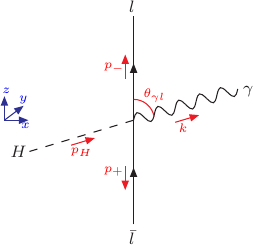}
\caption{Kinematical configuration for the $H(p_H)\to\gamma(k)l(p_-)\bar{l}(p_+)$ decay in the rest frame of the lepton-pair.}
\label{dilepRF}
\end{figure}

For the present computation corresponding to the $H(p_H)\to\gamma(k)l(p_-)\bar{l}(p_+)$ decay, the rest frame of the lepton-pair is chosen. Concretely, the $z$-axis is along the direction of the lepton, the $y$-axis is perpendicular to the decay plane and the photon has positive $x$-component, as represented in \figref{dilepRF}. The two independent kinematical variables are the dilepton invariant mass, $m_{\dilep}=\sqrt{s}$, and the angle between photon and lepton, $\thlep$.
The momentum of each particle is
\bear
&& \hspace{-5mm}p_H = \left(\sqrt{\mh^2+|\vec{k}|^2},|\vec{k}|\sin(\thlep),0,|\vec{k}|\cos(\thlep)\right) \,,\quad k = \left(|\vec{k}|,|\vec{k}|\sin(\thlep),0,|\vec{k}|\cos(\thlep)\right) \,,  \nn\\
&& \hspace{-5mm}p_- = \left(\sqrt{s}/2,0,0,\sqrt{s}\betal/2\right) \,,\quad p_+ = \left(\sqrt{s}/2,0,0,-\sqrt{s}\betal/2\right) \,\,\,\text{with}\,\,\,|\vec{k}|=\frac{\mh^2-s}{2\sqrt{s}} \,,
\eear
and $\betal=\sqrt{1-4\mlep^2/s}$ is the lepton velocity in this frame.

The two transverse polarization vectors of the photon, with the usual normalization, are
\be
\varepsilon_\pm(k) = \frac{1}{\sqrt{2}}\left(0,-i\cos(\thlep),\mp 1,i\sin(\thlep)\right)
\ee

Of course, we can chose the Higgs rest frame by performing the Lorentz transformation
\be
L^{(H)} = \begin{pmatrix}
\sqrt{\mh^2+|\vec{k}|^2}/\mh & -|\vec{k}|\sin(\thlep)/\mh & 0 & -|\vec{k}|\cos(\thlep)/\mh \\
0 & \cos(\thlep) & 0 & -\sin(\thlep) \\
0 & 0 & 1 & 0 \\
-|\vec{k}|/\mh & \sqrt{\mh^2+|\vec{k}|^2}\sin(\thlep)/\mh & 0 & \sqrt{\mh^2+|\vec{k}|^2}\cos(\thlep)/\mh \\
\end{pmatrix}\,.
\ee

In that frame, the photon momentum is 
\be
k^{(H)} =\left(\frac{\mh^2-s}{2\mh},0,0,\frac{\mh^2-s}{2\mh}\right) \,,
\ee
and a lower cut $E^\gamma_{cut}$ to the photon energy is imposed in order to avoid IR divergences in the tree level contribution. Then the upper bound on the dilepton invariant mass in \eqref{mdileprange} is obtained.

\section{Helicity amplitudes}
\label{App-helamps}

This Appendix collects the 8 helicity amplitudes and presents the numerical estimations for the decay width. The amplitudes, which were implemented in the numerical results of \secref{section-numresults}, include lepton mass effects and are written in terms of the generic 1-loop form factors.
Using the kinematics of \appref{App-kinem}, the amplitudes of \eqrefs{amptree}{amploop} as function of $s$ and $\thlep$ are:
{\allowdisplaybreaks
\bear
\mM_{+++} &=& \frac{-i\sqrt{2}\Ctree\kCP\sinthlep}{(\mh^2-s)(1-\costhlep^2\betal^2)} \left(\cosCP(\mh^2(1-\betal)+s(1+\betal)-8\mlep^2)+i\sinCP(\mh^2(1-\betal)-s(1+\betal))\right) \nn\\
&& -\frac{i\mlep(\mh^2-s)\sinthlep}{4\sqrt{2}}\left(\aone(1-\betal)+\atwo(1+\betal)+\bone(1+\betal)+\btwo(1-\betal)\right) \,,  \nn\\
\mM_{++-} &=& \frac{-i2\sqrt{2}\Ctree\kCP e^{i\dCP}(1+\costhlep)\mlep}{\sqrt{s}(1-\costhlep^2\betal^2)}  \nn\\
&& -\frac{i(\mh^2-s)(1+\costhlep)}{4\sqrt{2}\sqrt{s}}\left(2\aone\mlep^2+\atwo(s(1+\betal)-2\mlep^2)+\bone (s(1-\betal)-2\mlep^2)+2\btwo\mlep^2\right) \,,  \nn\\
\mM_{+-+} &=& \frac{-i2\sqrt{2}\Ctree\kCP e^{i\dCP}(1-\costhlep)\mlep}{\sqrt{s}(1-\costhlep^2\betal^2)}  \nn\\
&& -\frac{i(\mh^2-s)(1-\costhlep)}{4\sqrt{2}\sqrt{s}}\left(2\aone\mlep^2+\atwo(s(1-\betal)-2\mlep^2)+\bone (s(1+\betal)-2\mlep^2)+2\btwo\mlep^2\right) \,,  \nn\\
\mM_{+--} &=& \frac{-i\sqrt{2}\Ctree\kCP\sinthlep}{(\mh^2-s)(1-\costhlep^2\betal^2)} \left(\cosCP(\mh^2(1+\betal)+s(1-\betal)-8\mlep^2)+i\sinCP(\mh^2(1+\betal)-s(1-\betal))\right) \nn\\
&& -\frac{i\mlep(\mh^2-s)\sinthlep}{4\sqrt{2}}\left(\aone(1+\betal)+\atwo(1-\betal)+\bone(1-\betal)+\btwo(1+\betal)\right) \,,  \nn\\
\mM_{-++} &=& \frac{-i\sqrt{2}\Ctree\kCP\sinthlep}{(\mh^2-s)(1-\costhlep^2\betal^2)} \left(\cosCP(\mh^2(1+\betal)+s(1-\betal)-8\mlep^2)-i\sinCP(\mh^2(1+\betal)-s(1-\betal))\right) \nn\\
&& -\frac{i\mlep(\mh^2-s)\sinthlep}{4\sqrt{2}}\left(\aone(1-\betal)+\atwo(1+\betal)+\bone(1+\betal)+\btwo(1-\betal)\right) \,,  \nn\\
\mM_{-+-} &=& \frac{i2\sqrt{2}\Ctree\kCP e^{-i\dCP}(1-\costhlep)\mlep}{\sqrt{s}(1-\costhlep^2\betal^2)}  \nn\\
&& +\frac{i(\mh^2-s)(1-\costhlep)}{4\sqrt{2}\sqrt{s}}\left(\aone(s(1+\betal)-2\mlep^2)+2\atwo\mlep^2+2\bone\mlep^2+\btwo (s(1-\betal)-2\mlep^2)\right) \,,  \nn\\
\mM_{--+} &=& \frac{i2\sqrt{2}\Ctree\kCP e^{-i\dCP}(1+\costhlep)\mlep}{\sqrt{s}(1-\costhlep^2\betal^2)}  \nn\\
&& +\frac{i(\mh^2-s)(1+\costhlep)}{4\sqrt{2}\sqrt{s}}\left(\aone(s(1-\betal)-2\mlep^2)+2\atwo\mlep^2+2\bone\mlep^2+\btwo (s(1+\betal)-2\mlep^2)\right) \,,  \nn\\
\mM_{---} &=& \frac{-i\sqrt{2}\Ctree\kCP\sinthlep}{(\mh^2-s)(1-\costhlep^2\betal^2)} \left(\cosCP(\mh^2(1-\betal)+s(1+\betal)-8\mlep^2)-i\sinCP(\mh^2(1-\betal)-s(1+\betal))\right) \nn\\
&& -\frac{i\mlep(\mh^2-s)\sinthlep}{4\sqrt{2}}\left(\aone(1+\betal)+\atwo(1-\betal)+\bone(1-\betal)+\btwo(1+\betal)\right) \,.
\label{helamps}
\eear
}

For each helicity amplitude, the first and second lines correspond to tree level and 1-loop contributions, respectively. 
The form factors $a_{1,2}$ and $b_{1,2}$ do not depend on $\mlep$, then the complete dependence on lepton mass is explicitly shown in this equation. Of course, the SM is recovered for $\kCP=1$ and $\dCP=0$.

Some comments about interesting limits are in order. First, when photon has vanishing energy ($s=\mh^2$), the 1-loop contribution vanishes for all helicity amplitudes and the tree level yields to IR divergences for $\{+++,+--,-+++,---\}$ amplitudes (which are avoided by the lower cut $E_{cut}^\gamma$).
Secondly, when photon is collinear with lepton ($\thlep=0$) or with antilepton ($\thlep=\pi$), just two helicity amplitudes are non-vanishing for each case:

{\allowdisplaybreaks
\bear
&&\hspace{-8mm}\mM_{++-}\vert_{\thlep=0} = -\frac{i\sqrt{2}\Ctree\kCP e^{i\dCP}\sqrt{s}}{\mlep}  \nn\\
&& \hspace{18mm}-\frac{i(\mh^2-s)}{2\sqrt{2}\sqrt{s}}\left(2\aone\mlep^2+\atwo(s(1+\betal)-2\mlep^2)+\bone (s(1-\betal)-2\mlep^2)+2\btwo\mlep^2\right) \,,  \nn\\
&&\hspace{-8mm}\mM_{--+}\vert_{\thlep=0} = \frac{i\sqrt{2}\Ctree\kCP e^{-i\dCP}\sqrt{s}}{\mlep}  \nn\\
&& \hspace{18mm}+\frac{i(\mh^2-s)}{2\sqrt{2}\sqrt{s}}\left(\aone(s(1-\betal)-2\mlep^2)+2\atwo\mlep^2+2\bone\mlep^2+\btwo (s(1+\betal)-2\mlep^2)\right) \,,  \nn\\
&&\hspace{-8mm}\mM_{+-+}\vert_{\thlep=\pi} = -\frac{i\sqrt{2}\Ctree\kCP e^{i\dCP}\sqrt{s}}{\mlep}  \nn\\
&& \hspace{18mm}-\frac{i(\mh^2-s)}{2\sqrt{2}\sqrt{s}}\left(2\aone\mlep^2+\atwo(s(1-\betal)-2\mlep^2)+\bone (s(1+\betal)-2\mlep^2)+2\btwo\mlep^2\right) \,,  \nn\\
&&\hspace{-8mm}\mM_{-+-}\vert_{\thlep=\pi} = \frac{i\sqrt{2}\Ctree\kCP e^{-i\dCP}\sqrt{s}}{\mlep}  \nn\\
&& \hspace{18mm}+\frac{i(\mh^2-s)}{2\sqrt{2}\sqrt{s}}\left(\aone(s(1+\betal)-2\mlep^2)+2\atwo\mlep^2+2\bone\mlep^2+\btwo (s(1-\betal)-2\mlep^2)\right) \,.
\label{ampscol}
\eear
}

Thirdly, the massless lepton case ($\mlep=0$, $\betal=1$) results in the state of \eqref{psimlzero}. In particular, it also has IR divergences when photon is collinear with lepton and antilepton, as can be seen from \eqref{ampscol}.

\begin{table}[h!]
    \centering
\begin{tabular}{c||cc||c|c||c}
\hline
lepton case & $\sqrt{s}_{\rm min}$ [GeV] & $\sqrt{s}_{\rm max}$ [GeV] & $\Gamma_{\rm full}$ [KeV] & $\Gamma_{\rm hybrid}$ [KeV] & $N_{\rm Run\, 2+3}$  \\
\hline
tau-lepton & $12.5$ & $124$ & $31.07$ & $31.06$ & 63641 \\
 & $12.5$ & $30$ & $0.255$ & $0.138$ & 522 \\
\hline
muon & $12.5$ & $124$ & $0.932$ & $0.512$ & 1909 \\
 & $12.5$ & $30$ & $0.218$ & $0.027$ & 446 \\
\hline
electron & $12.5$ & $124$ & $0.556$ & $0.321$ & 1139 \\
 & $12.5$ & $30$ & $0.217$ & $0.026$ & 444 \\
\hline
\end{tabular}
\caption{Decay width $\Gamma_{H\to\gamma\dilep}$ for both full and hybrid computations, in the range of invariant masses $\in[\sqrt{s}_{\rm min},\sqrt{s}_{\rm max}]$. The estimation of the expected number of events for LHC Run 2 + Run 3 data is also included in the last column (see text for details).}
\label{tab:widths}
\end{table}

Furthermore, we focus on the dilepton invariant mass regime of \eqref{mdileprange} and no additional cuts over the final particles are applied in this work. The resulting decay width for the three flavors, using both full and hybrid computations, are presented in \tabref{tab:widths} (see also \figref{anatomyplots}). 
Two relevant energy ranges are considered in this table: a low-mass dilepton subsystem $\in[0.1\mh,30\,{\rm GeV}]$, as in~\cite{ATLAS:2021wwb}, and the complete range $\in[0.1\mh,\sqrt{s}_{\rm cut}]$.
These results are compatible with previous computations~\cite{Korchin:2014kha,Kachanovich:2020xyg,Aakvaag:2023xhy}.

The estimation of the expected number of events for LHC Run 2 + Run 3 data, using the full computation, is also included in the last column of \tabref{tab:widths}. In this estimation, the NNNLO Higgs boson production cross section at 13 TeV is 48.61 pb, the total decay width is 4.07 MeV and the luminosity of the combined Runs 2 and 3 is 350 fb$^{-1}$~\cite{Workman:2022ynf,Cepeda:2019klc}. In addition, it is assumed an identification efficiency of 0.7 for each lepton in final state and consider the inclusive decay channel for the tau-lepton. These results scale trivially with the lepton identification efficiency.
For the HL-LHC estimation, the expected cross section at 14 TeV is 54.67 pb and the luminosity is 3000 fb$^{-1}$, then we can expect 10 times more events. Furthermore, notice that this three-body decay channel is quite clean and the background originates predominantly from non-resonant $\dilep\gamma$ production~\cite{CMS:2018myz,ATLAS:2021wwb}.

\section{Main results using Dalitz plots representation}
\label{App-Dalitz}

Introducing the Mandelstam variables $s=(p_-+p_+)^2$, $t=(k+p_-)^2$ and $u=(k+p_+)^2$, we can plot the main results of \secref{section-numresults} using Dalitz plots representation, in order to be independent of the reference frame. 
These variables satisfy the relation $s+t+u=\mh^2+2\mlep^2$ and the angle of the photon and lepton is written as
\be
\cos(\thlep) = \frac{u-t}{(\mh^2-s)\sqrt{1-4\mlep^2/s}} \,,
\ee

\begin{figure}[h!]
    \centering
\begin{tabular}{ccr}
\includegraphics[width=0.28\textwidth]{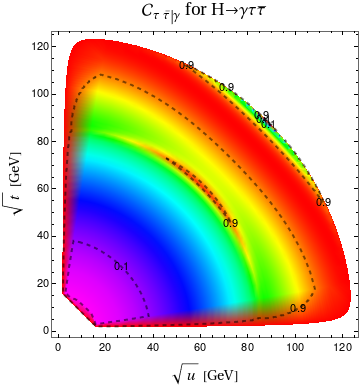} & \includegraphics[width=0.33\textwidth]{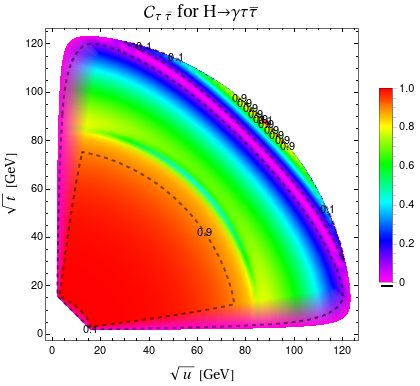} & \includegraphics[width=0.33\textwidth]{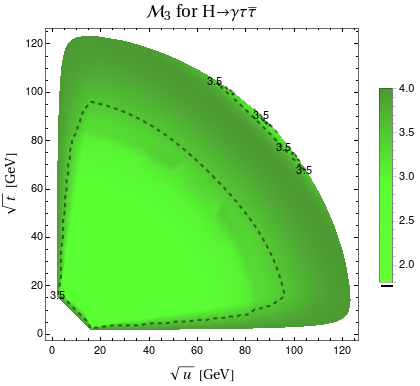} \\
\includegraphics[width=0.28\textwidth]{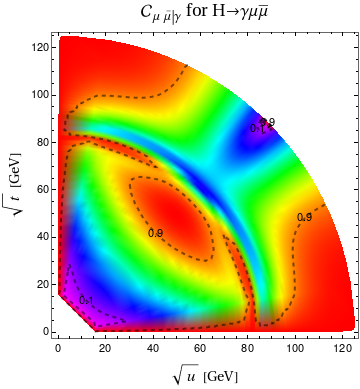} & \includegraphics[width=0.33\textwidth]{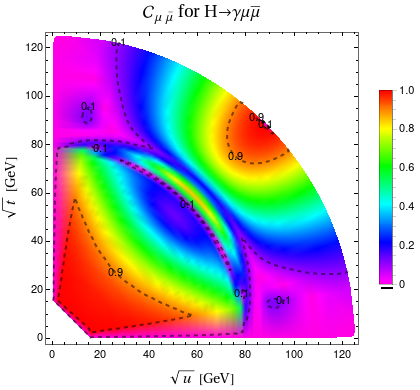} & \includegraphics[width=0.33\textwidth]{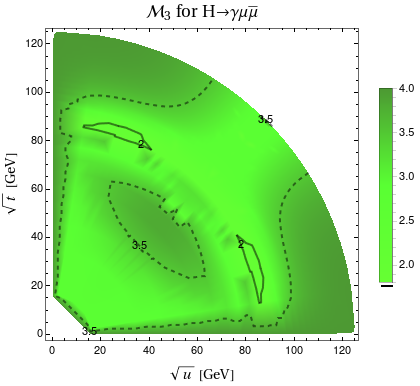} \\
\end{tabular}
\begin{tabular}{cccr}
\includegraphics[width=0.21\textwidth]{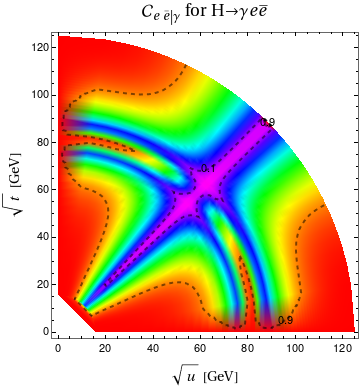} & \includegraphics[width=0.21\textwidth]{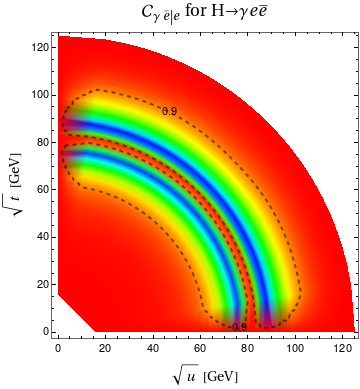} & \includegraphics[width=0.245\textwidth]{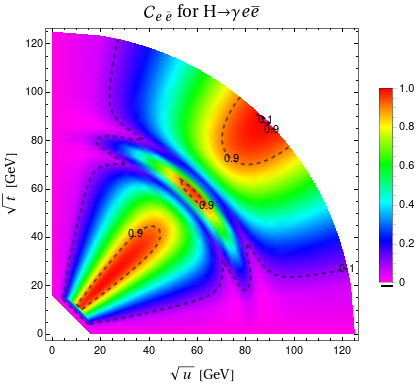} & \includegraphics[width=0.245\textwidth]{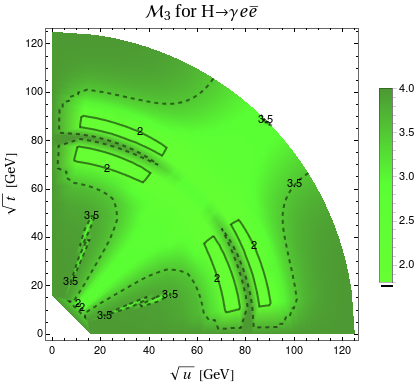} \\
\end{tabular}
\caption{Entanglement and non-locality quantifiers in the $[\sqrt{u},\sqrt{t}]$ plane corresponding to the full computation for tau-lepton (first row), muon (second row) and electron (third row).}
\label{Dalitz-plots}
\end{figure}

Hence \figref{Dalitz-plots} presents the results of the full computation in the $[\sqrt{u},\sqrt{t}]$ plane, already discussed in \figrefs{tau-plots}{el-plots}. This format may be more convenient from the experimental point of view in these three-body decays, however we chose the $[m_{\dilep},\cos(\thlep)]$ plane since is more intuitive for identifying kinematical configurations. Of course the physics behind these phenomena is the same in both cases.

\bibliography{TriEnt_Hdecays-arXiv_v3}

\end{document}